\def\versionno{ bi-branes  -- version 4.2  -- by kw --  15.3.07 }

\documentclass[12pt]{article}
\usepackage{latexsym,amsmath,amssymb,amsfonts,bbm,xypic}

\usepackage[T1]{fontenc}
\usepackage{amsthm}
\usepackage[all]{xy}
\usepackage{graphicx}

\usepackage{epstopdf,hyperref} 

\numberwithin{equation}{section}

\catcode`\@=11
\newif\if@fewtab\@fewtabtrue
{\count255=\time\divide\count255 by 60
\xdef\hourmin{\number\count255}
\multiply\count255 by-60\advance\count255 by\time
\xdef\hourmin{\hourmin:\ifnum\count255<10 0\fi\the\count255}}
\def\ps@draft{\let\@mkboth\@gobbletwo
    \def\@oddfoot{\hbox to 7 cm{\tiny \versionno
       \hfil}\hskip -7cm\hfil\rm\thepage \hfil {\tiny\draftdate}}
    \def\@oddhead{}
    \def\@evenhead{}\let\@evenfoot\@oddfoot}
\def\draftdate{\number\month/\number\day/\number\year\ \ \ \hourmin }

\catcode`\@=12

\def\C             {\mathbbm{C}}
\def\CP            {\mathbbm{C}\mathbb{P}}

\def\R             {\mathbbm{R}}

\def\RP            {\mathbbm{R}\mathbb{P}}
\def\Z             {\mathbb{Z}}

\def\id            {\mathrm{id}}

\def\BB            {F}  
\def\be            {\begin{equation}}
\def\bearl         {\begin{array}{l}}
\def\bearll        {\begin{array}{ll}}
\def\bfe           {{\bf1}}
\def\bti           {\,{\boxtimes}\,}

\def\calb          {\mathcal{B}}
\def\calc          {\mathcal{C}}
\def\calcft        {\mathfrak{C}}
\def\caln          {\mathcal{N}}
\def\cft           {conformal field theory}

\def\chii          {\raisebox{.11em}{$\chi$}}

\def\ee            {\end{equation}}
\def\eear          {\end{array}}
\def\eq            {\,{=}\,}
\newcommand\erf[1] {(\ref{#1})}
\newcommand\Frac[2]{\mbox{\large$\frac{#1}{#2}$}}
\def\g             {\mathfrak{g}}
\def\gbar          {\overline{\mathfrak{g}}}

\def\hy            {$\mbox{-\hspace{-.66 mm}-}$\linebreak[0]}
\def\Hy            {$\mbox{-\hspace{-.66 mm}-}$}
\def\iN            {\,{\in}\,}
\newcommand\labl[1]{\label{#1}\ee}
\def\lambdap       {{\lambda^+_{\phantom+}}}
\newcommand\lli[1]{ {}_{#1\!}}  
\newcommand\lui[1]{ {}^{#1\!}}  
\def\mult          {\mathrm{mult}}
\def\oti           {\,{\otimes}\,}
\def\rmd           {\mathrm{d}}

\def\Times         {\,{\times}\,}
\def\tr            {\mathrm{tr}}
\def\Ue            {\mathrm U(1)}

\makeatletter

\def\adress#1{\gdef\@adress{#1}}
\def\@adress{}
\def\preprint#1{\gdef\@preprint{#1}}
\def\@preprint{}
\def\@maketitle{
  \newpage
  \noindent
  \begin{tabular}{cc}
    \begin{minipage}[c]{0.4\textwidth}
      \begin{flushleft}
        \includegraphics[width=110pt]{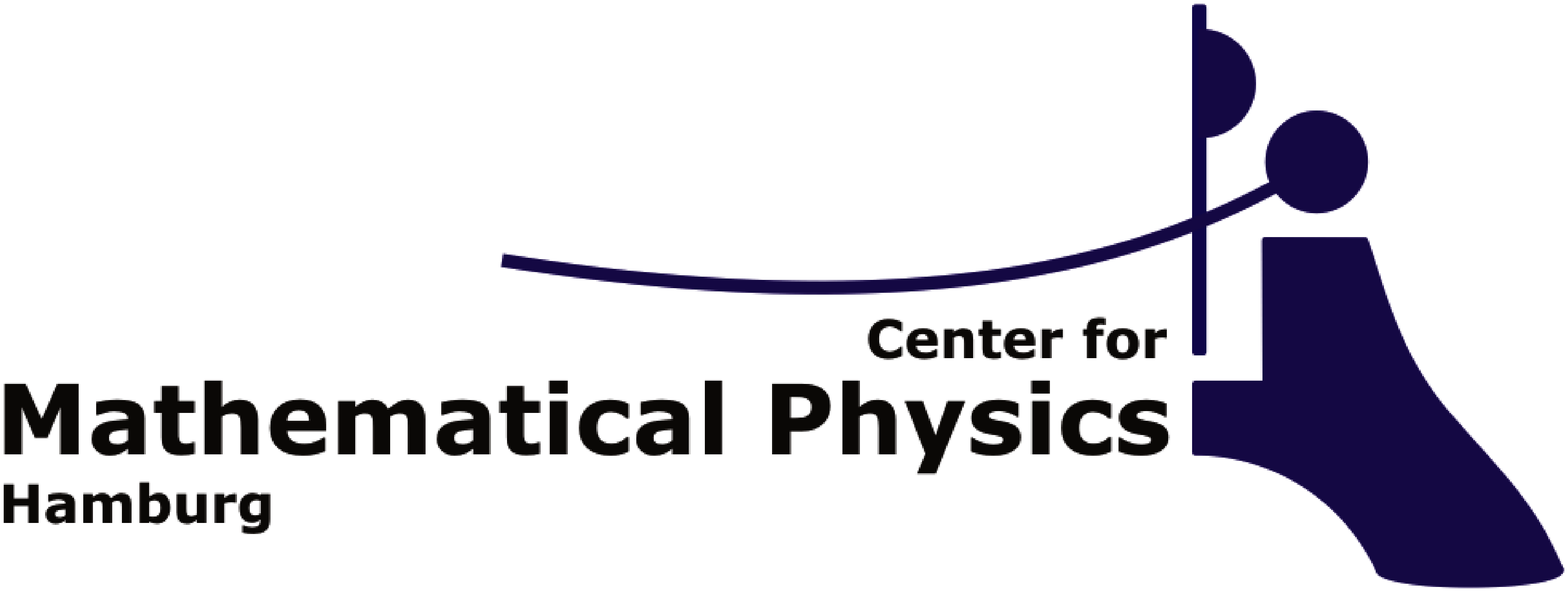}      
      \end{flushleft}
    \end{minipage}&
    \begin{minipage}[c]{0.55\textwidth}
      \begin{flushright}
      {\small\sf\@preprint}
      \end{flushright}
    \end{minipage}
  \end{tabular}
  \vskip 3cm
  \begin{center}
    \LARGE\@title
    \if!\@author!\else \vskip 1cm \large\@author\fi
    \if!\@adress!\else \vskip 1cm \normalsize\@adress\fi
  \end{center}
  \vskip 2cm
}

\newcommand{\alxydim}[2]{\begin{aligned}\xymatrix#1{#2}\end{aligned}}
\newcommand{\bbf}{\varpi}

\makeatother

\begin{document}


\title{
BI-BRANES:\, TARGET SPACE GEOMETRY \\[2pt] FOR WORLD SHEET TOPOLOGICAL DEFECTS}
\author{Jürgen Fuchs$\,^{1}$,~~ Christoph Schweigert$\,^2$,~~ Konrad
Waldorf$\,\,^2$}
\adress{$^{1}$ Teoretisk fysik\\Karlstads Universitet
\\Universitetsgatan 5\\ S\,--\,651 88 Karlstad \\ \bigskip
$^{2}$ Organisationseinheit Mathematik\\Schwerpunkt Algebra und
Zahlentheorie\\Universität
Hamburg\\Bundesstra\ss e 55\\D\,--\,20146 Hamburg }
\preprint{arxiv:hep-th/0703145\\Hamburger Beiträge zur Mathematik Nr.\
268\\ZMP-HH/07-04}
\maketitle

\thispagestyle{empty}

\begin{abstract}
\noindent
We establish that the relevant geometric data for the target space description 
of world sheet topological defects are submanifolds -- which we call bi-branes 
-- in the product $M_1\times M_2$ of the two target spaces involved. Very much 
like branes, they are equipped with a vector bundle, which in backgrounds with
non-trivial $B$-field is actually a twisted vector bundle.
We explain how to define Wess\hy Zumino terms in the presence of bi-branes 
and discuss the fusion of bi-branes. 
\\
In the case of WZW theories, symmetry preserving bi-branes are shown to
be biconjugacy classes. The algebra of functions on a biconjugacy class
is shown to be related, in the limit of large level, to the partition
function for defect fields. We finally indicate how the Verlinde
algebra arises in the fusion of WZW bi-branes.
\end{abstract}

\newpage

\tableofcontents

\thispagestyle{empty}

\newpage
\setcounter{page}{1}

\section{Introduction}

Sigma models have been one significant source of examples for two-dimensional
conformal field theories. 
They allow one to relate geometric structure on target space to field theoretic
quantities in the conformal field theory. This has has provided much insight,
not least for the interpretation of string theory. A particularly important 
observation has been the relationship between (conformal) world sheet boundary 
conditions and D-branes, which are, in their simplest incarnation, submanifolds 
of the target space equipped with a vector bundle. 

The target space of a sigma model has, at least, the structure of a 
(pseudo-)Riemannian manifold. Further structure on the target space is
introduced by the presence of the tachyon and of the antisymmetric Kalb-Ramond
$B$-field. While we will ignore the tachyon in the present article, we do take 
the $B$-field into account, wherever this is possible without rendering the 
exposition too technical. The appropriate geometric structure on target space 
needed to describe a non-trivial $B$-field background is a hermitian bundle 
gerbe, and for a D-brane the vector bundle gets replaced by a
twisted vector bundle, i.e.\ by a gerbe
module for the restriction of the gerbe to the world volume of the brane.

Apart from conformally invariant boundary conditions, two-dimensional
conformal field theories admit another, equally natural, structure:
topological defect lines. These objects are familiar from
statistical mechanics. Take, for example, the lattice version of the
Ising model: changing the coupling along all bonds that cross a specified line
from ferromagnetic to antiferromagnetic, produces a defect. Due to
the $\Z_2$-gauge invariance of the Ising model, the position of this defect can 
be moved around, as long as we do not cross the site of a spin that appears in 
the correlator of interest. If we do cross such a site, we are forced to change 
the sign of the spin variable. The defect thus comes with a well-defined 
rule for passing insertions in the bulk through the defect line.

Moreover, in the Ising model a pair of two such defect lines which run close 
to each other can be eliminated by a gauge
transformation; more generally, two defects can be joined to a single defect,
which gives rise to fusion rules between topological defects.
A similar phenomenon arises when we take boundary conditions into account: 
In the Ising model, a given boundary condition, say ``spin sup'',
combined with a parallel antiferromagnetic defect line can be replaced by
the boundary condition ``spin down''. More generally, there is a mixed fusion 
by which topological defects act on conformal boundary conditions.

Similarly as in the case of boundary conditions, in the CFT that is obtained
in the continuum limit this structure can be expected to result in defect 
lines along which correlation functions of bulk fields can have a branch-cut 
like behaviour. At least for rational conformal field theories, such defect 
lines appear naturally in algebraic approaches to CFT
\cite{pezu5,fuRs4}; in the TFT approach to RCFT correlators \cite{fuRs4}
a complete description of such defects is available \cite{ffrs3,ffrs5}. The 
TFT approach allows one, in particular, to compute the partition functions of 
bulk and boundary fields, and of defect fields (fields living on a defect line 
that can change the type of defect), as well as the fusion of two defects and of 
a defect with a conformal boundary condition.

More specifically, suppose a collection of conformal field theories is
{\em compatible\/} in the sense that they share a chiral symmetry algebra, 
including at least the Virasoro algebra. Note that in order for two conformal 
field theories to be compatible, they must in particular have the same Virasoro 
anomaly. A standard example of compatible theories are
the WZW models based on $SU(2)$ and on $SO(3)$ with the same value of the level.
We label the members of a compatible family of conformal field theories by 
indices $\{A_1,A_2,...\,\}$. There then exist (oriented) defects which
separate the conformal field theory of type $A_1$ present on a region of world
sheet to their left from a conformal field theory of type $A_2$ to their right 
hand side. Such a topological defect will be denoted by $\lli{A_1}B_{A_2}$. Then
the fusion of defects associates to two defects $\lli{A_1}B_{A_2}$ and 
$\lli{A_2}B_{A_3}$ a defect of type $\lli{A_1}B_{A_3}$:
  \be
  \lli{A_1}B_{A_3} = \lli{A_1}B_{A_2} \star_{\!A_2} \lli{A_2}B_{A_3} \,.
  \ee
The second type of fusion associates to a defect $\lli{A_1}B_{A_2}$ and 
boundary condition $\lli{A_2}N$ for the theory of type $A_2$ a boundary 
condition $\lli{A_1}N$ for the theory of type $A_1$,
  \be
  \lli{A_1}N = \lli{A_1}B_{A_2} \star_{\!A_2} \lli{A_2}N \,.
  \ee
In the framework of \cite{fuRs4,ffrs5}, the labels $\{A_1,A_2,...\,\}$
correspond to certain algebras in the representation category of the chiral 
symmetry algebra. These algebras encode in particular the partition functions, 
including a modular invariant bulk partition function and partition functions 
for boundary and defect fields. Branes are described by modules, and defects 
by bimodules, of these algebras; the fusion operation $\star_{\!A}$ is realized 
as the tensor product over $A$. 

It has also been understood \cite{ffrs3,ffrs5} that topological defects encode 
information both on internal symmetries and on dualities of a conformal field 
theory; this includes in particular T-dualities.

\medskip

In view of the relevance of target space structures for string theoretic 
interpretations, it is natural to ask whether a target space description exists 
for conformal defects as well. The answer to this question is the primary result
of the present paper.

Suppose we are given two compatible conformal field theories, corresponding to
target spaces $M_1$ and $M_2$, respectively. We show that conformal defects 
correspond to submanifolds of the product $M_1\Times M_2$. Furthermore,
very much in the same way as for a brane, this submanifold has to be
endowed with a vector bundle (again, in the presence of a non-trivial $B$-field
this is a twisted vector bundle). For theories based on current
algebras -- compactified free bosons and Wess\hy Zumino\hy Witten theories --
we study the relevant submanifolds in detail. For simplicity, in this paper
we restrict our attention to the cases of a single compactified free boson 
and of the WZW model based on a compact, connected and simply connected Lie 
group. It is clear, however, that when combined with standard techniques 
developed for $D$-branes, the concepts presented here allow one to extend our 
results to more general classes of conformal field theories, in particular to 
WZW theories on non-simply connected groups, coset theories, theories of 
several free bosons compactified on a torus, and orbifolds of such theories. 

\medskip

In the rest of this paper we will proceed as follows. Inspired by the 
calculation of the scattering of closed string states in the presence of 
D-branes \cite{dfpslr,fffs}, in Section \ref{sec2} we analyze
scattering processes in the presence of defect lines, considering theories 
with current symmetries and defects of type $\lli AB_A$. In these cases we 
have $M_1 \eq M_2 \eq M$, and the target space $M$ is a compact connected 
Lie group. In the simply connected case the relevant submanifold of 
$M\Times M$ turns out to be a biconjugacy class, i.e.\ is of the form
  \be
  \mathcal{B}_{h_1,h_2} := \left\lbrace (g_1^{},g_2^{})\iN G \Times G \;|\;
  \exists\, x,y\iN G\text{: }g_1^{}\eq x^{}h_1^{}y^{-1}\text{,}\,
  g_2^{}\eq x h_2^{} y^{-1} \right\rbrace .
  \ee
This is analogous to the role played by conjugacy classes 
\cite{alsc2,fffs,staN5} in the description of boundary conditions.
Correspondingly, the so-called 2-characters 
  \be
  \begin{array}{lrcl} \chii^{(2)}_\lambda:& G \Times G &\!\to\!& \C \\
              & (g_1,g_2) &\!\mapsto\!& \tr_{H_\lambda}^{}(g_1g_2) 
  \eear \ee
take the role that characters play in the theory of branes. We will therefore
refer to the target space objects that describe defects as {\em bi-branes\/}.

It should be appreciated that while the multiplication of the Lie group $G$
enters in the specific form of bi-branes for WZW theories, the description of 
defects in general does {\em not\/} require a multiplication on target space. 
Rather, the relevant structure for bi-branes separating theories with target 
spaces $M_1$ and $M_2$ are suitable submanifolds of $M_1\Times M_2$.

\medskip

In Section \ref{sec3} we discuss the intrinsic geometry of biconjugacy
classes and relate the algebra of functions on a biconjugacy class to
the algebra of defect fields; we can then exhibit a two-form on the
biconjugacy class that trivializes the difference of the three-form
field strengths on the two backgrounds involved. In Section \ref{secWZ} we
show how these data can be employed to construct a Wess\hy Zumino term 
in situations in which the topology of both the target space and the bi-brane 
are particularly simple; a proof that the so constructed Wess\hy Zumino term
is well-defined, as well as the description of the Wess\hy Zumino term for more
general target spaces and/or bi-branes, are relegated to appendices. Finally,
Section \ref{sec4} is devoted to aspects 
of the fusion of two bi-branes and of the fusion of a bi-brane to a brane;
we provide in particular an argument for how the Verlinde algebra arises 
as the fusion algebra of symmetry preserving bi-branes on simply connected 
Lie groups. A short outlook is supplied in Section \ref{sec5}.


\section{Scattering of bulk fields in the backgrounds of defects}
\label{sec2}

One rationale for assigning a target space geometry to a conformal field theory
is to study the scattering of bulk fields. This is based on the general idea 
(see e.g.\ \cite{frgA}) that (a subspace of) the space of bulk fields can be
identified with a truncation and deformation of the algebra of functions on the 
target space. In the case of branes this amounts, in tree level approximation
to string theory scattering amplitudes, to computing the two-point functions of 
bulk fields on a disk with given boundary condition. By factorization
to a three-point function on the sphere and a one-point function on
the disk, this can be reduced \cite{dfpslr,fffs}
to the computation of one-point functions of bulk fields on the disk.

Here we are interested in probing the target space geometry for
a topological defect $B$ on the world sheet, again using the scattering of
bulk fields. In tree-level approximation we have to consider the two-point
functions of bulk fields on a world sheet that is a sphere $S^2$ containing
a closed defect line $B$. Without loss of generality, we can take the defect 
line to be along the equator of the sphere. If both bulk field insertions are 
on the same hemisphere, then by factorization we just obtain
the correlator in the absence of a defect, multiplied by the quantum
dimension of the defect \cite{fjfrs}. To get information on the
relevant geometry of the target space, we must thus consider the situation
with the two bulk insertions on different hemispheres, i.e.\ on different
sides of the defect line.

\medskip

For theories with current symmetry we will use the following notation.
By $\gbar$ we denote a finite-dimensional reductive complex Lie algebra.
Special cases of particular interest are those where $\gbar$ is simple,
and the abelian Lie algebra $\mathfrak u(1)\,{\oplus}\cdots
{\oplus}\,\mathfrak u(1)$. By $G$ we denote the simplest compact Lie group
with Lie algebra (the compact real form of) $\gbar$. Thus for semisimple 
$\gbar$, $G$ is the connected simply connected compact Lie group with Lie 
algebra $\gbar$, while for reductive Lie algebras we take in addition the 
direct product with $d$ copies of $U(1)$, with $d$ the dimension of the center 
of $\gbar$. For concreteness, the reader might wish to keep in mind the two 
special cases $\gbar\eq \mathfrak u(1)$ and $\gbar\eq \mathfrak{su}(2)$, 
with $G \eq U(1)$ and $G \eq SU(2)$, respectively.

By $\g$ we denote the nontrivial central extension of the 
loop algebra of $\gbar$; if $\gbar$ is simple, $\g$ is an
untwisted affine Lie algebra, while for $\gbar$ abelian we have a direct sum
of Heisenberg algebras with identified centers.
We fix the value of the level $k$ for each simple ideal of $\gbar$; 
the irreducible highest weight representations are then 
classified by the set $P_k$ of dominant integral weights $\lambda$ at level 
$k$. Analogously the irreducible finite-dimensional representations of 
$\gbar$ are labeled by the set $P$ of dominant integral $\gbar$-weights.
In particular, for $\gbar\eq \mathfrak u(1)$, Fock spaces are labeled by 
momentum, so that $P_k \eq P \eq \R$, while for $\gbar\eq \mathfrak{su}(2)$,
at positive integral level $k$ the relevant sets are
$P_k \eq \{0,1,...\,,k\}$ and $P \eq \Z_{\ge0}$.

Thus for any $\lambda\iN P$ we have a finite-dimensional $\gbar$-module
$H_\lambda$ (for $\gbar\eq \mathfrak{su}(2)$ its dimension is $\lambda{+}1$).
We may as well regard $H_\lambda$ as a $G$-module; its character is
  \be
  \begin{array}{lrcl}
  \chii_\lambda:& G &\!\to\!& \C^\times \\
  & g & \!\mapsto\!& \mathrm{tr}_{H_\lambda}^{} R_\lambda(g) \,.
  \eear \ee
Via taking the horizontal part of an affine weight, we can regard $P_k$ as a
subset of $P$. The irreducible $\g$-module with highest weight $\lambda\iN P_k$ 
is infinite-dimensional, with finite-dimensional homogeneous subspaces; we 
identify its zero-grade subspace with the finite-dimensional 
$\gbar$-module $H_\lambda$. Finally, by $\lambda^+$ we denote the highest weight
of the representation that is conjugate to $H_\lambda$. For $\gbar\eq \mathfrak 
u(1)$, this is the representation with opposite $\mathfrak u(1)$-charge; 
for $\gbar\eq \mathfrak{su}(2)$, every representation is self-conjugate.

\medskip

Returning to our preceding discussion, we now consider the correlation function
on $S^2$ of two bulk fields labeled by
$\mathfrak{g}\,{\oplus}\,\mathfrak{g}$-modules $H_\lambda^{}\bti H_\lambdap$
and $H_\mu\bti H_{\mu^+}$ inserted, respectively, at
the north and south pole of $S^2$, with a defect $B$ along the equator.
Further, we restrict our attention to the the so-called Cardy case, in which
the bulk partition function is given by charge conjugation, boundary
conditions are labeled by primary fields and the annulus coefficients
are fusion rules \cite{card9}. In the Cardy case also the topological defects 
are labeled by the same set $P_k$ as the left- and right-moving parts of the 
bulk fields. In the sequel we abbreviate the defect
$B\eq B_\alpha$ with $\alpha\iN P_k$ by $\alpha$.

By holomorphic factorization, any correlator on $S^2$ is an element
of the space of conformal blocks on the double cover of $S^2$, which
consists of the disjoint union of two copies of $\CP^1$ with
opposite orientation. For the correlator $\mathcal D_{\alpha;\lambda\mu}$ of 
two bulk fields on $S^2$ with a defect line $\alpha$, we thus deal with
a four-point block $D_{\lambda\mu}$ on $\CP^1 \,{\sqcup}\, \CP^1$,
which is an element of the algebraic dual of the tensor product vector
space $H_\lambda \oti H_{\lambda^+_{}} \oti H_\mu \oti H_{\mu^+}$.
Similarly as in \cite{fffs} we consider the particular correlator
  \begin{equation}
  \label{1}
  \mathcal G^{abcd}_{\alpha;\lambda\mu}(v\oti \tilde v\oti w\oti\tilde w)
  := \mathcal{D}_{\alpha;\lambda\mu} ( J^a_{-1} v \oti J^b_{-1} \tilde v
  \oti J^c_{-1} w \oti J^d_{-1} \tilde w) \,,
  \end{equation} 
where by $J^a_n$, with $a$ a labeling a basis of $\gbar$, we denote
the modes of the currents $J^a(z)$ (for the corresponding basis elements of
$\gbar$ we write $\bar J^a$).

In order for the correlator (\ref{1}) to be non-zero we need $\mu\eq\lambda^+$.
The states $v$ and $\tilde w$ are then vectors in the $\gbar$-module $H_\lambda$,
while $\tilde v$ and $w$ are states in the $\gbar$-module $H_{\lambda^+_{}}$,
with these $\gbar$-modules regarded as the zero-grade
subspaces of the corresponding $\mathfrak{g}$-modules.

\medskip

To determine the correlation function \erf1, we first study the four-point
conformal blocks $D_{\lambda\lambdap}$ on $\CP^1 \,{\sqcup}\, \CP^1$.
They decompose into a tensor product of two-point blocks on the two copies of 
$\CP^1$, $D_{\lambda\lambdap} \eq \BB_{\lambda}^{} \oti \BB_\lambdap$.
The chiral Ward identities for left and right movers read
  \be
  D_{\lambda\lambdap} \circ \big( J^a_{-n} \oti \bfe \oti \bfe \oti \bfe
  + \bfe \oti \bfe \oti J^a_n \oti \bfe \big) = 0
  \ee
and
  \be
  D_{\lambda\lambdap} \circ \big( \bfe \oti J^a_{-n} \oti \bfe \oti \bfe
  + \bfe \oti \bfe \oti \bfe \oti J^a_n \big) = 0 \,,
  \ee
respectively, for all $a\eq1,2,...\,,\dim(\gbar)$ and all $n\iN\Z$.
Together with the highest weight properties of
$w$ and $\tilde w$ and with the commutation relations of $\g$, the
Ward identities imply
  \be
  \hspace{-.3cm} \begin{array}{ll}
  D_{\lambda\lambdap}
  (J^a_{-1}v\oti J^b_{-1}\tilde v \oti J^c_{-1}w\oti J^d_{-1}\tilde w)
  \!\!\! & = D_{\lambda\lambdap} (v\oti \tilde v\oti J^a_1 J^c_{-1}w\oti
  J^b_1 J^d_{-1}\tilde w)
  \\{}\\[-.7em]&
  = \BB_{\lambda}^{} (v\oti [J^a_1,J^c_{-1}]w)\, \BB_\lambdap
  (\tilde v\oti [J^b_1,J^d_{-1}]\tilde w)
  \\{}\\[-.7em]&
  = \big[ \BB_\lambda^{}(v\oti
  [\bar J^a,\bar J^c]w) + k\,\kappa^{ac} \BB_\lambda^{}(v\oti w) \big]
  \\{}\\[-.9em] &\hspace*{1.7cm}
  \cdot\, \big[ \BB_\lambdap(\tilde v\oti[\bar J^b,\bar J^d]\tilde w)
  + k\,\kappa^{bd} \BB_\lambdap(\tilde v\oti\tilde w) \big] .
  \eear \labl{e24}

We expect that a direct contact to the geometry of compact Lie groups exists in 
the weak coupling limit, i.e.\ in the limit of large level $k$.
Accordingly we only keep those terms in \erf{e24}
which are of leading order in $k$; they are proportional to the Killing form 
of $\gbar$ and correspond to graviton and dilaton scattering; if $\gbar$ is 
abelian, they are the only terms present. In this limit we obtain the expression
  \be
  k^2 \kappa^{ac} \kappa^{bd}\, \BB_\lambda^{}(v\oti w)\, \BB_\lambdap
  (\tilde v\oti\tilde w)
  =: k^2 \kappa^{ac} \kappa^{bd}\, D_{\lambda\lambdap}^\infty
  (v\oti \tilde v\oti w\oti\tilde w)\,\text{.}
  \labl3

As in \cite{fffs}, at this point we invoke the Peter\hy Weyl theorem, so as to 
identify the space $\bigoplus_{\lambda\in P_k}\! H_\lambda\bti H_{\lambda^+_{}}$
with a subspace of the space $\mathcal{F}(G)$ of functions on the Lie group $G$.
This way, equation \erf{3} allows us to associate to a defect a
linear function on $\mathcal{F}(G)$, i.e.\ a distribution.
Before computing this distribution, which essentially amounts to a Fourier 
transformation, we notice that while boundary conditions give a distribution 
on $G$, defects give a distribution on the product manifold $G\Times G$. As a 
consequence, defects will be associated to submanifolds of $G\Times G$. This 
also fits nicely with the philosophy behind the so-called folding trick
\cite{woaf}, by which a conformal defect separating two conformal field theories
$\calcft_{A_1}$ and $\calcft_{A_2}$ with the same conformal anomaly is related 
to a conformally invariant boundary condition in the product theory
$\calcft_{A_1} \Times \overline{\calcft_{A_2}}$. It should be kept in mind, 
however, that in this article we are only concerned with topological
defects, which constitute a specific subclass of conformal defects.

\medskip

Let us now Fourier transform the result \erf3 according to the rules of 
\cite{fffs}, to obtain a distribution on $G\Times G$. We first note that
the Fourier transformation of a linear form $D$ on the space 
$\bigoplus_{\lambda,\mu\in P} H_\lambda \bti H_{\lambda^+_{}}
\bti H_\mu \bti H_{\mu^+_{}}$ reads
  \begin{eqnarray}
  &\hspace{-1.3cm}D(v\oti \tilde v\oti w\oti \tilde w) \!\!
  &= \int_{G\times G} \mathrm{d}g\,\mathrm{d}g'\, \widetilde D(g,g')^*
  \sum_{\lambda,\mu\in P}
  \left \langle \tilde v\oti \tilde w \right | R_\lambda(g) \oti
  R_\mu(g') \left | v\oti w \right \rangle
  \nonumber  \\&&=
  \int_{G} \rmd g\, \sum_{\lambda\in P} \left \langle \tilde v
  \right | R_\lambda(g) \left | v \right \rangle
  \int_{G} \rmd g'\, \sum_{\mu\in P} \left \langle \tilde w \right | R_\mu(g')
  \left | w \right \rangle \widetilde D(g,g')^* ,
  \end{eqnarray} 
and that its inverse is given by
  \be
  \displaystyle \widetilde D(g,g') = \sum_{\mu_1;i,j} \sum_{\mu_2;k,l}
  N_{\mu_1}N_{\mu_2}\, D(v_i\oti\tilde v_j\oti w_k\oti\tilde w_l)
  \, \left \langle \tilde v_j \right | R_{\mu_1}(g) \left| v_i \right
  \rangle
  \left \langle \tilde w_l \right | R_{\mu_2}(g') \left | w_k \right
  \rangle\text{,}
  \ee
with $\{v_i\}$ a basis of $H_{\mu_1}$ and $\{\tilde v_i\}$ the dual basis 
of $H_{\mu_1^+}$, and analogously for for $w_k$ and $\tilde w_k$.
Here the normalization factors $N_{\mu_i}$ are given by
$N_{\mu} \eq \sqrt{ |H_\mu|/ |G| }$
with $|H_\mu|$ the dimension of $H_\mu$ and $|G|$ the volume of $G$.\,%
  \footnote{~Note that, like e.g.\ in \cite{bads,fuwu}, we do not take
  the volume of $G$ to be normalized to 1. Rather, the `physical' radius of $G$
  should be $\sqrt{k\alpha'}$, i.e. $|G|$ is proportional to
  $(k\alpha')^{\dim(G)/2}$.}

For the functions (\ref{3}) of our interest this prescription yields
  \begin{eqnarray}
  &\widetilde D_{\lambda\lambdap}^\infty(g,g') \!\!
  &= \sum_{\mu_1^{},\mu_2^{}\in P}\! N_{\mu_1^{}}N_{\mu_2^{}} \sum_{i,j,k,l}
  \left \langle \tilde v_j \right | R_{\mu_1^{}}(g) \left | v_i \right \rangle
  \BB_\lambda^{}(v_i \oti v_k) \, \left \langle \tilde v_l \right |
  R_{\mu_2^{}}(g') \left | v_k \right \rangle \BB_\lambdap
  (\tilde v_j \oti \tilde v_l)
  \nonumber \\
  && = \, N_\lambda^2 \sum_{i,j,k,l}
  \left \langle \tilde v_j \right | R_\lambda^{}(g) \left | v_i \right \rangle
  \BB_\lambda^{}(v_i \oti v_k)
  \, \left \langle \tilde v_l \right | R_\lambdap(g') \left| v_k
  \right\rangle \BB_\lambdap (\tilde v_j \oti \tilde v_l)\text{.}
  \end{eqnarray}
By the identities $R_{\lambda^+_{}}(g) \eq (R_\lambda(g^{-1}))^{\mathrm{t}}$, 
where the superscript indicates the transpose matrix, and
$\BB_\lambda(v_i \oti v_k) \eq \delta_{i,k}$, this reduces to
  \be
  \widetilde D_{\lambda\lambdap}^\infty(g,g')
  = N_\lambda^2 \sum_{i,j}
  \big( R_\lambda(g) \big)^j_{i} \big( R_\lambda(g'^{-1}) \big)^i_{j}
  = N_\lambda^2 \, \chii_\lambda^{}(gg'^{-1}) \,\mathrm{.}
  \ee

Here 2-characters of $G$ pop up. 2-characters are functions on the
Cartesian product $G\Times G$ of a group with itself. They
first appeared in \cite{frob7} in the expansion of group
determinants. As compared to characters, they contain more information
about the group than characters; e.g.\ in contrast to characters, they
allow one to determine whether a representation is real or pseudo-real.
(Still, 2-characters and characters do not determine a group up to
isomorphism. A surprisingly recent result \cite{hoJo} states that a
group is determined by its 1-, 2- and 3-characters.)

\medskip

Next we use the results of the TFT approach (following the lines of Section 4 
of \cite{fuRs10}) to express 
the correlation functions in terms of conformal blocks: we have
  \be
  \mathcal{D}_{\alpha;\lambda\mu} = \frac{S_{\lambda,\alpha}}{S_{0,\lambda}}
  \, D_{\lambda\lambdap}
  = \chii_\alpha^{}(h_\lambda)^*_{} D_{\lambda\lambdap}
  = \frac{S_{0,\alpha}}{S_{0,\lambda}}\,
  \chii_\lambda^{}(h_\alpha)^*_{} D_{\lambda\lambdap} \,,
  \labl4
where similarly as in \cite{fffs} we introduced the group element
  \be
  h_\alpha := \exp(2\pi\mathrm{i}\, \hat y_\alpha) \,,
  \ee
with $\hat y_\alpha$ the Cartan subalgebra element dual to the weight
  \be
  y_\alpha := \frac{\alpha+\rho}{k+g^{\vee}} \,\in \gbar_0^* \,\mathrm{.}
  \ee
($\rho$ denotes the Weyl vector and $g^\vee$ the dual Coxeter
number of $\gbar$.) For the sum
  \be
  \mathcal G^{abcd}_{\alpha} := \sum_{\lambda\in P_k}
  \mathcal G^{abcd}_{\alpha;\lambda\lambdap}
  \labl{sumG}
of two-point correlators, which is the analogue of a boundary state,
we thus obtain, at large $k$,
  \be
  \widetilde G^{abcd}_{\alpha}(g,g')
  = k^2\kappa^{ac}\kappa^{bd} \sum_{\lambda\in P_k} N_\lambda^2\,
  \frac{S_{0,\alpha}}{S_{0,\lambda}}\, \chii_\lambda^{}(h_\alpha)^*_{}\,
  \chii_\lambda^{}(gg'^{-1}) \,\text{.}
  \ee
Furthermore, using that at large $k$ the quantum dimension 
$S_{0,\lambda}/S_{0,0}$ approaches the ordinary dimension $|H_\lambda|$ and 
$P_k$ can be replaced by $P$, this reduces to
  \be
  \widetilde G^{abcd}_{\alpha}(g,g') = k^2\kappa^{ac}\kappa^{bd}\,
  \frac{|H_\alpha|}{|G|} \sum_{\lambda\in P}
  \chii_\lambda^{}(h_\alpha)^*_{}\,\chii_\lambda^{}(gg'^{-1}) \,\text{.}
  \ee
Up to normalization this is a delta distribution on the conjugacy class
$\mathcal{C}_\alpha \,{\equiv}\, \mathcal{C}_{h_\alpha}^{}$ of $G$:
  \be
  \sum_{\lambda\in P} \chii_\lambda^{}(h_\alpha)^*_{}
  \chii_\lambda^{}(gg'^{-1}) = \Frac{|G|}{|\mathcal{C}_\alpha|}\,
  \delta_{\mathcal{C}_\alpha}(gg'^{-1}) \, \text{.}
  \ee
Thus we finally arrive at
  \be
  \widetilde G^{abcd}_{\alpha}(g,g') \,
  \stackrel{k\to\infty}{{-}\!\!\!\longrightarrow}\, k^2\kappa^{ac}\kappa^{bd}\,
  \Frac{|H_\alpha|} {|\mathcal{C}_\alpha|}\,
  \delta_{\mathcal{C}_\alpha}(gg'^{-1}) \, \text{.}
  \ee
In short, for given topological defect $\alpha$, in the large level limit the 
analogue \erf{sumG} of the boundary state is concentrated on those pairs 
$(g,g')\iN G\Times G$ whose product $gg'^{-1}$ lies in $\mathcal{C}_\alpha$.


\section{The world volume of WZW bi-branes} \label{sec3}

\subsection{Biconjugacy classes}

According to the scattering calculation in the previous section, the geometric 
object in $G\Times G$ that is relevant for the description of a defect $\alpha$ 
is the set of those points points $(g_1,g_2)$ of $G \Times G$ such that $g_1^{}
g_2^{-1}$ lies in the conjugacy class $\mathcal{C}_{\alpha}$ of $G$. These 
subsets of $G\Times G$ are actually submanifolds; we wish to describe them
in more detail. To this end we introduce the following notion: For a compact 
connected Lie group $G$ and elements $h_1,h_2\iN G$ we call the submanifold
  \be
  \mathcal{B}_{h_1,h_2} := \left\lbrace (g_1^{},g_2^{})\iN G \Times G \;|\;
  \exists\, x_1^{},x_2^{}\iN G\text{: }g_1^{}\eq x_1^{}h_1^{}x_2^{-1}\text{,}\,
  g_2^{}\eq x_1^{} h_2^{} x_2^{-1} \right\rbrace 
  \,\subset G\Times G
  \ee
the {\em biconjugacy class\/} of the pair $(h_1,h_2)$.

\medskip

Biconjugacy classes inherit from the diagonal left and diagonal right actions
of $G$ on $G\Times G$ two commuting actions of $G$. For the defects we are 
describing here, these two $G$-actions correspond to the two independent 
preserved current symmetries.

Obviously, 2-characters are constant on biconjugacy classes. In fact, very 
much like the characters of irreducible $G$-representations form a natural basis 
for the functions on the space of conjugacy classes, the 2-characters
of irreducible representations form a basis
for the space of functions on biconjugacy classes.

Next we observe that the smooth map
  \be
  \begin{array}{lrcl}
  \tilde \mu : & G \Times G &\to& G\\& (g_1,g_2) &\!\mapsto\!& g_1^{}g_2^{-1}
  \eear
  \labl{mudef}
intertwines the diagonal left and diagonal right action of
$G$ on $G\Times G$ and the adjoint and trivial actions of $G$ on itself,
respectively. Put differently, $\tilde\mu$ defines the structure of
a trivializable $G$-equivariant principal $G$-bundle over $G$. Indeed,
the $G$-action on the fibers is by diagonal right multiplication, so that the
$G$-equivariant diffeomorphism $t{:}\ (g_1,g_2) \,{\mapsto}\, (g_1g_2,g_2)$
furnishes a global trivialization, where the trivial $G$-bundle 
$p_1{:}\ G \Times G \,{\to}\, G$ over $G$ projects on the first component.

\medskip

It now follows that a biconjugacy class in $G \Times G$ is the preimage of 
a conjugacy class in $G$ under the  projection $\tilde \mu$ defined in 
\erf{mudef}:
  \be
  \mathcal{B}_{h_1,h_2}^{} = \tilde\mu^{-1}(\mathcal{C}_{h_1^{}h_2^{-1}}) =
  \big\lbrace (g_1,g_2) \iN G \Times G \;|\; g_1^{}g_2^{-1}
  \iN \mathcal{C}_{h_1^{}h_2^{-1}} \big\rbrace \,\text{;}
  \labl{preim}
in particular,
  \be
  \mathcal{B}_{h_1,h_2}^{} = \mathcal{B}_{h_1^{}h_2^{-1},e} \,\text{.}
  \ee
To establish the relation \erf{preim},
we observe that for every element $(g_1,g_2) \iN\mathcal{B}_{h_1,h_2}$
we have $g_1^{} \eq x_1^{}h_1^{}x_2^{-1}$ and $g_2^{} \eq x_1^{}h_2^{}x_2^{-1}$ 
for some $x_1^{}.x_2^{} \iN G$, and hence $g_1^{}g_2^{-1} \eq 
x_1^{}h_1^{}h_2^{-1}x_1^{-1} \iN \mathcal{C}_{h_1^{}h_2^{-1}}$. Conversely, 
given $(g_1^{},g_2^{})\iN G \Times G$ such that there exists some $x\iN G$ with 
$xg_1^{}g_2^{-1}x^{-1} \eq h_1^{}h_2^{-1}$, we set
$x_1^{} \,{:=}\, x^{-1}$ and $x_2^{} \,{:=}\, g_2^{-1}x^{-1}h_2^{}$ and obtain 
$g_1^{} \eq x_1^{}h_1^{}x_2^{-1}$ and $g_2^{} \eq x_1^{}h_2^{}x_2^{-1}$, which 
shows that $(g_1,g_2) \iN \mathcal{B}_{h_1,h_2}$.

\medskip

To conclude, biconjugacy classes have the topology of a direct product of $G$ 
with a conjugacy class. Thus for simply connected groups, they are in particular
simply connected. The scattering of closed string states in WZW theories detects
bi-branes corresponding to biconjugacy classes for which $h_1^{}h_2^{-1}$ is a 
regular element of $G$; this closely parallels
the findings of \cite{fffs} for branes.


\subsection{World volume quantization}

As further evidence for the relation between biconjugacy classes and WZW 
defects, we will now establish that the defect fields associated to a
topological defect furnish a quantization of the space of functions on a 
biconjugacy class. 
Note that besides bulk fields there also exist other types of fields in the
presence of defects \cite{ffrs5}: disorder fields, at which defect lines start 
or end, and defect fields, which live on a defect line and can change the type 
of the defect. There is a distinguished type of defect, acting as a unit with
respect to fusion, called the invisible defect. Across this defect, every bulk 
fields is smooth. Disorder fields are in fact special defect fields: those 
changing the invisible defect to some other defect or vice versa. Similarly, 
bulk fields can be regarded as defect fields preserving the invisible defect 
and thus as special disorder fields.

Since there are two commuting actions of $G$ on the world volume of a 
biconjugacy class, the space $\mathcal{F}(\mathcal{B}_{h_1,h_2})$ has the
structure of a $G\Times G$-module.
This can be compared with the situation for conjugacy classes, which describe 
WZW branes. A conjugacy class $\mathcal C$ carries a natural $G$-action, 
the adjoint action, which turns the space of $\mathcal{F}(\mathcal C)$ of
functions on $\mathcal C$ into a $G$-module. As pointed out in \cite{fffs},
only regular conjugacy classes are relevant to the situation of our interest.
A regular conjugacy class is isomorphic to $G/T$, with $T$ a maximal torus of
$G$, and there is an isomorphism
  \be
  \mathcal{F}(G/T) \,\cong\, \bigoplus_{\lambda\in P} \mult_\lambda(0) \,
  H_\lambda
  \labl{FGT}
of $G$-modules, where $\mult_\lambda(0)$ denotes the multiplicity of the weight
$0$ in the highest weight $\gbar$-module $H_\lambda$.

This $G$-module structure is related, in the large-level limit,
to the $G$-module structure of a subset of the space of boundary fields for
the corresponding WZW brane. Note that in the present context we should perform
the large-$k$ limit in a way such that the geometric conjugacy class is kept 
fixed. As a consequence, the weight labeling the boundary condition depends on
the level. More specifically, just like in \cite{fffs} we must consider weights
$\alpha \eq \alpha(k)$ such that
  \be
  \alpha_0 := \frac{\alpha(k) +\rho}{k+g^\vee}
  \ee
is constant. The large-$k$ limit of the WZW annulus coefficients 
$\mathrm A_{\lambda \alpha}^{\,\beta}$ for the case of simply connected $G$ 
reads \cite{fffs}
  \be
  \lim_{k\to\infty}^{}\lui{(k)}\mathrm A_{\lambda\, \alpha(k)}^{\,\,\beta(k)}
  = \delta_{\alpha_0,\beta_0}\, \mult_\lambda(0) \, .
  \ee
This result can be interpreted as follows. In the large-level limit,
only open strings starting and ending at the same brane survive. As a
$G$-module, they have the algebra of functions on the brane as a limit;
this substantiates the idea that the space of open strings constitutes a 
quantization of the world volume of the brane.

For bi-branes, we can obtain an analogous result by using $G\Times G$-modules
in place of $G$-modules. To describe the intrinsic geometry of the
bi-brane $\mathcal{B}_{h_1,h_2}$, with $h_1$ and $h_2$ regular elements
of $G$, we first note that the bijection
  \be \begin{array}{lrcl}
  (p_1\Times\tilde\mu): & \mathcal{B}_{h_1,h_2} &\!\to\!& G \Times
  \calc_{h_1^{}h_2^{-1}} \\{}\\[-.9em]
  &(g_1^{},g_2^{}) &\!\mapsto\!& (g_1^{},g_1^{}g_2^{-1})
  \eear \ee
intertwines two pairs of $G$-actions: first, the diagonal left action of $G$ on
$\mathcal{B}_{h_1,h_2}$, i.e.\ $\rho(h)((g_1,g_2)) \eq (hg_1,hg_2)$, is 
intertwined with $G$ acting from the left on itself and by the adjoint action on 
$\calc_{h_1^{}h_2^{-1}}$; and second, the diagonal right action on $\mathcal{B}
_{h_1,h_2}$ is intertwined with the right action on $G$ and the trivial action 
on $\calc_{h_1^{}h_2^{-1}}$. The $G\Times G$-module structure of the space of 
functions on $\mathcal{B}_{h_1,h_2}$ now follows easily; we have
  \be
  \mathcal{F}(\mathcal{B}_{h_1,h_2})
  \,\cong\, \mathcal{F}(G \Times \calc_{h_1^{}h_2^{-1}})
  \,\cong\, \mathcal{F}(G) \otimes \mathcal{F}(\calc_{h_1^{}h_2^{-1}}) \,.
  \ee
Further, by the Peter\hy Weyl theorem we have $\mathcal{F}(G)\,{\cong}\,
\bigoplus_{\mu\in P}\! H_\mu^{}\bti H_{\mu^+_{}}$, while the 
$G$-module structure of $\mathcal{F}(\calc_{h_1^{}h_2^{-1}})$ is given by 
\erf{FGT}. Thus after decomposing the tensor product we obtain 
  \be
  \mathcal{F}(\mathcal{B}) \,\cong\, \bigoplus_{\lambda,\mu\in P}
  \Big( \sum_{\nu\in P}^{}\, \overline\caln_{\!\nu\mu^+_{}}^{\,\,\lambda}\,
  \mult_\nu(0) \Big)\, H_\lambda \bti H_{\mu} \,,
  \labl{FB}
where $\overline\caln_{\!\nu\mu^+_{}}^{\,\,\lambda}$ is the multiplicity of the 
irreducible $\gbar$-module $H_\lambda$ in the tensor product 
$H_\nu\oti H_{\mu^+_{}}$.

The decomposition \erf{FB} has to be compared with the multiplicities 
$Z_{\mu\nu}^{\alpha\beta}$ for defect fields with chiral labels $\mu,\nu$ that
change a defect $\alpha$ to a defect $\beta$. A simple calculation 
in the TFT approach to rational conformal field theories (compare Section 5.10
of \cite{fuRs4}) shows that, in the Cardy case, this multiplicity is an 
ordinary fusion rule. Accordingly, we have at level $k$
  \be
  \lui{(k)} Z_{\lambda\mu}^{\alpha(k)\,\beta(k)}
  = \lui{(k)}\caln_{\lambda\,\mu\,\alpha(k)}^{\,\,\beta(k)}
  \equiv \sum_{\nu\in P_k} \lui{(k)}\caln_{\lambda\mu}^{\,\nu} \,
  \lui{(k)}\caln_{\nu\,\alpha(k)}^{\,\,\beta(k)} \, . 
  \ee
The large-$k$ limit of the two factors in this result follows easily: the fusion
rules $\lui{(k)}\caln_{\lambda\mu}^{\,\nu}$
tend to tensor product multiplicities, while the limit of the second 
factor is the same as the one computed above for the annulus coefficients 
(which for the Cardy case coincide with ordinary fusion rules). Thus we find
  \be
  \lim_{k\to\infty}^{} \lui{(k)} Z_{\lambda\mu}^{\alpha(k)\,\beta(k)}
  = \delta_{\alpha_0,\beta_0}\sum_{\nu\in P}
  \overline\caln_{\!\lambda\mu}^{\,\,\nu} \, \mult_{\nu}(0) 
  = \delta_{\alpha_0,\beta_0}\sum_{\nu\in P}
  \overline\caln_{\!\nu\mu^+_{}}^{\,\,\lambda}\, \mult_{\nu}(0) \,,
  \ee
where in the second equality the charge conjugation properties of the tensor
product multiplicities are used. This is in full agreement with the 
$G\Times G$-module structure \erf{FB} of the space $\mathcal{F}(\mathcal{B})$ 
of functions on the bi-brane. Analogously as for branes, this substantiates 
the idea that the algebra of defect fields can be regarded as a quantization 
of the space of functions on the bi-brane.


\subsection{Trivialization of the $H$-field}

As is well-known \cite{witt8}, conformal invariance for theories with
non-abelian currents requires a non-trivial $B$-field background. While the
$B$-field is defined only locally, its curvature $H$ is a globally defined
three-form. One important property of branes is the fact that the
restriction of $H$ to the corresponding submanifolds is exact. 
For symmetric branes in the WZW model based on $\mathfrak g$ at level $k$,
the curvature is the three-form
  \be
  H = \Frac{k}{6} \left\langle \theta
  \wedge [\theta \wedge \theta] \, \right\rangle\text{,}
  \ee
where we have denoted by $\theta$ the left-invariant Maurer\hy Cartan form
on $G$, which is a $\gbar$-valued
one-form, and by $\langle\cdot\,,\cdot\rangle$ the Killing form on $\gbar$.
Restricted to a conjugacy class $\mathcal{C}_h$, the three-form $H$ can be
written as the derivative of a $G$-invariant two-form $\omega_h$,
  \be
  H|_{\mathcal{C}_h}^{} = \mathrm{d}\omega_h \,\text{.}
  \labl{Htriv}
We will now see that bi-branes have properties that generalize this behaviour.

Consider again the map $\tilde\mu$ whose restriction maps the bi-brane 
$\mathcal{B}_{h_1,h_2}$ to the conjugacy class $\calc_{h_1^{}h_2^{-1}}$.  
We introduce the two-form 
  \be
  \bbf_{h_1^{},h_2^{}}:= \tilde\mu^* \omega_{h_1^{}h_2^{-1}} -
  \Frac{k}{2}\,\langle p_1^*\theta\wedge p_2^*\theta\rangle
\label{3.15}
  \ee
on $\mathcal{B}_{h_1,h_2}$, where $p_i$, $i\eq1,2$, is the projection from 
$G\Times G\to G$ on its $i$th factor, and 
both summands are restricted to the submanifold $\mathcal{B}_{h_1,h_2}$
of $G\Times G$. From the intertwining properties of $\tilde\mu$ it follows
that the two-form $\bbf$ is bi-invariant. Analogously to the equality 
(\ref{Htriv}) on the world volume of a brane, on the world volume 
$\mathcal{B}_{h_1,h_2}$ of the bi-brane the identity 
  \be
  p_1^* H = p_2^* H + \rmd \bbf_{h_1,h_2}
  \label{2triv} \ee
holds; in other words: on $\mathcal{B}_{h_1,h_2}$, the difference of the
$H$-fields of the two target spaces involved is exact and equals the derivative 
of the 
two-form $\bbf_{h_1,h_2}$.

To establish the identity \erf{2triv}, we first recall the relation
  \be
  \tilde\mu^*H = p_1^*H - p_2^*H +\Frac{k}{2}\, \rmd \langle p_1^*\theta\wedge
  p_2^*\theta\rangle 
  \ee
(compare e.g.\ the proof of proposition 3.2 of \cite{almm})
which in the derivation of the Polyakov\hy Wiegmann formula accounts for
the correct behaviour of the Wess\hy Zumino term. On the other hand, we find
  \be
  \left(\tilde\mu^*H\right)\!|_{\mathcal{B}_{h_1,h_2}}^{}
  = \tilde\mu^*(H|_{\calc_{h_1^{}h_2^{-1}}}^{})
  = \tilde\mu^*(\rmd\omega_{h_1^{}h_2^{-1}} )
  = \rmd \tilde\mu^* \omega_{h_1^{}h_2^{-1}} \, ;
  \ee
together with the definition of $\bbf_{h_1,h_2}$
the last two equations imply \erf{2triv}.

\medskip

At this point it is worth mentioning the notion of a quasi-Hamiltonian
$G$-space which has been introduced in \cite{almm}. As shown 
in \cite{almm}, both conjugacy classes and the ``double'' $G\Times G$ are 
examples of such spaces. However, the reader should be warned that, while the 
case of conjugacy classes is directly relevant for the discussion of branes, 
the double as considered in \cite{almm} is endowed with a $G\Times G$-action
that does not restrict to the bi-brane submanifolds.


\section{The Wess\Hy Zumino term in the presence of defects} \label{secWZ}

Having identified a two-form $\bbf$ on the bi-brane that trivializes 
the restriction of the difference of the $H$-fields, we are in a position to
study the Wess\hy Zumino term for situations with particularly simple topology.
The analysis closely parallels the one in \cite{fist8}. As in the case of 
branes, a general and more satisfactory analysis must be based on the notion 
of hermitian bundle 
gerbes. A first discussion of these issues can be found in Appendix \ref{appB}.

To attain a situation with sufficiently simple topology, we restrict our 
attention in the sequel to 2-connected target spaces $M_1$ and $M_2$, i.e.\ 
besides being connected and simply connected, the manifolds $M_i$ also satisfy 
$\pi_2(M_i)\eq0$ (this includes in particular compact connected and
simply connected simple Lie groups). Because a bundle gerbe over a 2-connected 
space is completely determined by its curvature, which is a closed three-form
with integral periods, we may then consider target spaces $M_1$ and $M_2$
with closed integral three-forms $H_1$ and $H_2$. 

A similar phenomenon occurs for bi-branes if we make the additional assumption 
that the world volume of a bi-brane is connected and simply connected:
the two-form $\bbf$ that trivializes the difference of the three-forms is a 
sufficient substitute for the structure that is needed  in the general case as 
described in Appendix \ref{appB}. Note that all these assumptions are in 
particular met for WZW bi-branes of simply connected compact Lie groups.

Under these assumptions, we arrive at the following simplified definition of 
a bi-brane: A 
{\em simply connected $M_1$-$M_2$-bi-brane\/} between 2-connected target spaces
$M_1$ and $M_2$ with three-forms $H_i \iN \Omega^3(M_i)$, $i\eq1,2$, is a 
simply connected submanifold $Q$ of $M_1 \Times M_2$ together with
a two-form $\bbf\iN\Omega^2(Q)$ such that
  \be
  p_1^* H|_Q = p_2^* H|_Q + \rmd \bbf \,\text{.}
  \labl{4.1}

\medskip

The classical Wess\hy Zumino\hy Witten model is a theory of maps from
a two-dimensional world sheet to a target space. The space of maps has to be
chosen in a way conforming with the correlator of interest. For example, for
world sheets with non-empty boundary it is required that the boundary of
the world sheet is mapped into the world volume of a WZW brane.
Here our aim is to describe correlators with defect lines. We merely consider 
the simplest situation: a closed oriented world sheet $\Sigma$ with an embedded 
oriented circle $S\,{\subset}\, \Sigma$ that separates the world sheet into two 
components, $\Sigma \eq \Sigma_1 \,{\cup_{S}}\, \Sigma_2$, which we assume to 
inherit the orientation of $\Sigma$. Without loss of generality we assume 
$\partial\Sigma_1 \eq S$ and 
$\partial \Sigma_2 \eq \overline{S}$ as equalities of oriented manifolds,
where $\overline{S}$ is the manifold $S$ with opposite orientation.

We assume that the defect separates regions that support conformally invariant 
sigma models with target spaces $M_1$ and $M_2$ and consider pairs of maps
  \be
  \phi_i:\quad \Sigma_i \to M_i
  \ee
such that the image of the combined map
  \be \begin{array}{lrcl}
  \phi_S:& S &\!\to\!& M_1 \Times M_2 \\ & s &\!\mapsto\!& (\phi_1(s),\phi_2(s))
  \eear \ee
takes its values in the submanifold $Q$.

\medskip

We next wish to find the Wess\hy Zumino part of the action. First, since 
$Q$ is simply connected, there exists a two-dimensional oriented submanifold $D$
of $Q$ with $\partial D\eq\phi_S(S)$. We can glue the images of this disk under 
the projections $p_i{:}\ M_1 \Times M_2 \to M_i$ along their boundaries on the 
images $\phi_i(\Sigma_i)$ of the
the world sheets, and obtain two-dimensional oriented closed
submanifolds. Because we have required $\pi_2(M_i)\eq 0$, we can fill those to 
three-dimensional oriented submanifolds $B_i \,{\subset}\, M_i$ such that
  \be
  \partial B_1 = \phi_1(\Sigma_1) \,{\cup}\, p_1(\overline{D})
  \quad~\text{ and }\quad
  \partial B_2=\phi_2(\Sigma_2) \,{\cup}\, p_2(D) \,\text{.}
  \labl{neu2}
Equipped with such choices of submanifolds, we define
  \be
  S[\phi_1,\phi_2] := \int_{\!B_1}\! H_1^{}
  + \int_{\!B_2}\!H_2^{} + \int_{\!D}\bbf \,\text{.}
  \labl{neu1}
Note that superficially the expression \erf{neu1} depends  on the choices of
the manifolds $B_1$, $B_2$ and $D$. However, the ambiguities are integers, so 
that the exponential of \erf{neu1} is actually well-defined. This can be shown 
with the help of a homology theory based on two manifolds $M_1$ and $M_2$ and 
a submanifold $Q\,{\subset}\, M_1 \Times M_2$, which we set up in Appendix 
\ref{appA}. For the dual cohomology 
theory a theorem of de Rham type holds; it allows us to express a cohomology 
class with values in $\R$ as a triple of differential forms. The triple 
$(H_1,H_2,\bbf)$ then furnishes an example of a cocycle in this cohomology 
theory. As we show in Appendix \ref{appA}, the ambiguities of \erf{neu1} arise 
as the pairing of the cohomology class of $(H_1,H_2,\bbf)$ with a cycle in 
homology that results from different choices of the submanifolds $D$, $B_1$ and 
$B_2$. We then show that if the cocycle $(H_1,H_2,\bbf)$ corresponds to a 
cohomology class with values in $\Z$ -- we shall call such a
triple integral -- the ambiguities of \erf{neu1} are integers.

This is analogous to the
discussion of the Wess\hy Zumino term in the presence of branes \cite{fist8}:
in that case the relative cohomology of the pair $(M,Q)$ is relevant, where $Q$
is the world volume of the brane. The three-form $H$ and the 2-form $\omega$
on $Q$ define a cocycle in the relative cohomology with values in $\R$, and the 
Wess\hy Zumino term is the pairing of $(H,\omega)$ with a certain cycle. Its 
well-definedness imposes the condition that $(H,\omega)$ is integral, i.e.\ lies
in the cohomology with values in $\Z$. As in the case of branes, the 
integrality condition described above imposes 
severe restrictions on the biconjugacy classes that can describe defect 
lines. In fact, only those biconjugacy classes qualify which are of the form 
$\tilde\mu^{-1}(\calc)$, where $\calc\,{\subset}\, G$ is a suitable conjugacy 
class, namely one that supports a gerbe module which leads to a boundary 
condition preserving all chiral currents at level $k$. It should be appreciated,
though, that the two-form on the biconjugacy class differs from the pull-back 
of the two-form on the conjugacy class, and in fact there is no sensible way in 
which a gerbe bimodule can be seen as the pull-back of a gerbe module.

\medskip

In Appendix \ref{appB} we show how one can drop the restrictions
$\pi_2(M_i)\eq \pi_1(M_i)\eq 0$ on the topology of the background and
$\pi_1(Q)\eq 0$ on the topology of the bi-brane world volume. In the absence of
these conditions, it is not enough any longer to work with the two-form $\bbf$ 
on the bi-brane and the curvature three-forms $H_i$ on the backgrounds. Rather, 
connection-type data must be taken into account. This can be achieved using 
hermitian bundle gerbes, together with a new notion to be introduced
in Appendix \ref{appB}: {\em gerbe bimodules\/}. We refer to the same appendix 
for the definition of a Wess\hy Zu\-mino term in this general situation.
To show that the proposed Wess\hy Zumino term restores the conformal
symmetry of correlators with defects is beyond the scope of this article.


\section{Fusion of bi-branes} \label{sec4}

As pointed out in the introduction, there are two natural notions of fusion 
involving bi-branes: the fusion of two bi-branes, and the fusion of a bi-brane 
and a brane to a brane. In both cases, the fusion of elementary (bi-)branes 
yields, in general, a superposition of elementary (bi-)branes.

As has been seen in the algebraic approach, for WZW defects that preserve
all current symmetries there exists a natural notion of duality. It can be
characterized by the property that the fusion of a bi-brane and its dual
contains the special bi-brane which with respect to fusion acts as the identity.
Ignoring the shift in the location of bi-branes by the Weyl vector,
this is the bi-brane whose world volume is the biconjugacy class
$\mathcal{B}_{(e,e)}$, i.e.\ the diagonal $G\,{\subset}\,G\Times G$.
Upon quantization, the functions on this special bi-brane are related to
ordinary bulk fields, rather than general defect fields.

By invoking this duality, instead of working with the fusion rules 
  \be
  \calb_\alpha \star \calb_\beta
  = \sum_\gamma \caln_{\alpha\beta}^{\;\gamma}\, \calb_\gamma
  \ee
of bi-branes we sometimes consider the multiplicities
  \be
  \caln_{\alpha\beta\gamma}:= \caln_{\alpha\beta}^{\;\gamma^\vee} . 
  \ee
These structure constants are, in general, not symmetric; from the results of
the algebraic approach, however, we expect them to be invariant under cyclic 
permutations. The algebraic approach also predicts that in the case of compact 
connected and simply connected Lie groups, the constants
$\caln_{\alpha\beta}^{\;\gamma}$ are just the ordinary fusion multiplicities 
arising in the chiral theory, which satisfy the Verlinde formula.


\subsection{World volume fusion}

We first consider the effect of fusion on world volumes. In this context, the 
notation becomes more transparent when considering at once bi-branes describing 
defects that separate different target spaces $M_1$ and $M_2$.

The action of correspondences on sheaves suggests to consider the following
prescription: For the fusion of an $M_1$-$M_2$-bi-brane with world volume 
$B\,{\subseteq}\, M_1\Times M_2$ and an $M_2$-brane with world volume 
$V \,{\subseteq}\, M_2$ one should consider
  \be
  B\star V := p_1^{}\!\left( B \,{\cap}\, p_2^{-1}(V) \right)
  \labl{WxV}
with $p_i$ the $i$th projection $M_1\Times M_2\to M_i$. In general 
$B\,{\star}\,V$ is only a subset, rather than a sub\-manifold, of $M_1$. On a 
heuristic level one would expect, however, that the quantization of the branes 
\cite{bads} selects a finite superposition of branes, which
then should reproduce the results obtained in the TFT approach. 
The quantization conditions on the positions of branes require
additional geometric structure on the branes, namely twisted
vector bundles, and involve a subtle interplay of this structure with the 
background $B$-field. We will exhibit in examples how the required 
finite superposition of branes or bi-branes arises after geometric quantization.

Similarly, the fusion of an $M_1$-$M_2$-bi-brane $B$ with an
$M_2$-$M_3$-bi-brane $B'$ uses projections $p_{ij}$ from the triple product
$M_1\Times M_2\Times M_3$ to the two-fold products $M_i\Times M_j$:
  \be
  B \star B' := p_{13}^{}\big( p_{12}^{-1}(B)\,{\cap}\,p_{23}^{-1}(B') \big) \,.
  \labl{WxW}
Again the question of quantization should be addressed. This issue turns out to
be largely parallel to what happens in the mixed fusion of bi-branes to branes, 
and accordingly we will concentrate on the case of mixed fusion.


\subsection{Bi-branes of the compactified free boson at fixed radius}

We consider a free boson compactified on a circle $S_R^1$ of radius $R$ and
restrict ourselves, for the moment, to defects separating two world sheet
regions that support one and the same theory. In this
situation, it does not harm to identify the circle with the Lie
group $\Ue\,{\cong}\, \{ z\iN\C \,|\, |z|\eq1 \}$.

We consider two types of branes: D0-branes $V^{(0)}_x$ are localized
at the position $x\iN\R\bmod2\pi R\Z$. D1-branes, in contrast, wrap the whole 
circle. The D1-brane characterized by a Wilson line
$\alpha\iN \R\bmod\frac1{2\pi R}\,\Z$ will be denoted by $V^{(1)}_\alpha$;
the Wilson line describes a flat connection on $S^1_R$.

The world volume of a bi-brane on $S_R^1$ is a submanifold of 
$S_R^1\Times S_R^1$ of the form
  \be
  B_{x} := \{ (y,y{-}x) \,|\, y \iN \R\bmod2\pi R\Z \}
  \ee
with $x\iN \R\bmod2\pi R\Z$. $B_{x}$ has the topology of a circle, and according
to our general considerations in Appendix \ref{sec:BG} it must be endowed with a
flat connection, i.e.\ with a Wilson line $\alpha$. As a consequence, the 
natural parameters for bi-branes of a compactified free boson are a pair 
$(x,\alpha)$ taking
values in two dual circles describing a position on $S^1$ and a Wilson line.
We will write $\calb_{(x,\alpha)} \,{\equiv}\, (B_x,\alpha)$ for such bi-branes.

For the fusion of a bi-brane $\calb_{(x,\alpha)}$ and a D0-brane $V^{(0)}_y$ 
we have
  \be\begin{array}{rl}
  p_2^{-1}(V^{(0)}_y) = \{ (y',y) \,|\, y'\iN[0,2\pi R) \} \,, \quad&
  B_x \cap p_2^{-1}(V^{(0)}_y) = \{ (x{+}y,y) \} 
  \\{}\\[-.7em] \mathrm{and} \quad&
  p_1^{}\big( B_x \,{\cap}\, p_2^{-1}(V^{(0)}_y) \big) = \{ x{+}y \} \,,
  \eear\ee
so that the prescription \erf{WxV} yields
  \be
  \calb_{(x,\alpha)} \star V^{(0)}_y = V^{(0)}_{x+y} \,.
  \ee
Thus the fusion with a defect of type $\calb_{(x,\alpha)}$ acts on D0-branes
as a translation by $x$ in position space.

For the fusion of a bi-brane $\calb_{(x,\alpha)}$ and a D1-brane $V^{(1)}_\beta$,
we need to take the flat line bundle on the bi-brane into account. We first pull
back the line bundle on $V^{(1)}_\beta$ along the projection $p_2$ to a line 
bundle on $S^1_R\Times S^1_R$; then we restrict it to the world volume $B_x$ of 
$\calb_{(x,\alpha)}$ and tensor this restriction
with the line bundle on $\calb_{(x,\alpha)}$ described by the Wilson
line $\alpha$. This gives a line bundle with Wilson line $\alpha{+}\beta$
on the world volume of the bi-brane that can be pushed down along the projection
$p_1$ to a line bundle with the same Wilson line on $S^1_R$. We conclude that
  \be
  \calb_{(x,\alpha)} \star V^{(1)}_\beta = V^{(1)}_{\alpha+\beta} \,.
  \ee
Thus the fusion with a defect of type $\calb_{(x,\alpha)}$ acts on D1-branes
as a translation by $\alpha$ in the space of Wilson lines.

We can similarly compute the fusion of two bi-branes $\calb_{(x,\alpha)}$
and $\calb_{(x',\alpha')}$: we have
  \be\bearl
  p_{12}^{-1}(B_x) = \{ (y,y{-}x,y') \,|\, y,y'\iN[0,2\pi R) \} \,,
  \\{}\\[-.7em]
  p_{23}^{-1}(B_{x'}) = \{ (y,y',y'{-}x') \,|\, y,y'\iN[0,2\pi R) \} \,,
  \\{}\\[-.7em]
  p_{13}^{}\big( p_{12}^{-1}(B_x) \,{\cap}\, p_{23}^{-1}(B_{x'}) \big)
  = \{ (y,y{-}x{-}x') \,|\, y\iN[0,2\pi R) \} \,,
  \eear\ee
so that the position variables of bi-branes add up under fusion. To understand 
the behaviour of Wilson lines, we take into account the flat line bundles by 
pulling them back to $S_R^1\Times S_R^1\Times S_R^1$ and tensoring them. Then 
as in the case of mixed fusion, the Wilson lines add up. We thus obtain
  \be
  \calb_{(x_1,\alpha_1)}\star \calb_{(x_2,\alpha_2)}
  = \calb_{(x_1+x_2,\alpha_1+\alpha_2)} \,.
  \ee
Hence we find that both the position and Wilson line variable of bi-branes add 
up under fusion. This result exactly matches the fusion of the first set of 
defects that are derived algebraically in \cite{fgrS}; for these both the 
left- and right-moving currents are preserved, $J_1(z)\eq J_2(z)$ and 
$\bar J_1(\bar z)\eq \bar J_2(\bar z)$, for $z$ a point on the defect line.
One can also consider the case that 
one or both of the currents are only preserved up to a non-trivial automorphism;
the $\mathfrak u(1)$ current algebra has only a single non-trivial automorphism,
acting as $J\,{\mapsto}\,{-}J$. The simplest case then turns out to be that both
$J_1(z)\eq{-}J_2(z)$ and $\bar J_1(\bar z)\eq{-}\bar J_2(\bar z)$; in this case 
one obtains submanifolds of the form $B \eq \{(y\bmod 2\pi R\Z,h{-}y\bmod 
2\pi R\Z) \,|\, y\iN\R\}$. The case of different automorphisms for left movers 
and right movers is more subtle; we expect the corresponding bi-branes to fill
the whole product space. Also, formula \erf{3.15} suggests that the two-form on 
the bi-brane should be proportional to $\pm\, \rmd\theta_1\wedge \rmd\theta_2$, 
with the sign depending on the chirality on which the non-trivial automorphism 
acts. These issues will not be addressed in the present paper.


\subsection{Bi-branes for the compactified free boson at different radii}

We next turn our attention to bi-branes which describe topological defects that 
separate a region which supports a boson compactified on a circle of radius 
$R_1$ from a region supporting a boson compactified at radius $R_2$. We describe
the product space by two coordinates $x_1$ and $x_2$, with $x_i$ to be taken
modulo $2\pi R_i \Z$. The bi-brane world volumes are
  \be
  B_h := \{ (y\bmod 2\pi R_1\Z\, , y{-}h\bmod 2\pi R_2\Z) \,|\,  y\iN \R \} \,.
  \ee
If the ratio $R_1/R_2$ is not rational, this set is isomorphic to $\R$ and
fills $S^1_{R_1}\Times S^1_{R_2}$ densely. Accordingly there
are no Wilson line variables. The algebraic approach shows that in
this situation there is a single defect that preserves all current
symmetries \cite{fgrS}; in particular, $h$ is not a physical parameter.

We thus assume that the ratio of the two radii is rational,
  \be
  {R_1} / {R_2} = r / s
  \ee
with $r,s$ coprime positive integers. The bi-brane world volume then has 
length $2\pi sR_1\eq 2\pi rR_2$ and admits a Wilson line variable, to be 
taken modulo $1/(2\pi sR_1) \eq 1/(2\pi rR_2)$. It wraps $s$ times in 
$R_1$-direction; hence the geometric parameter, when measured on the 
$x_2$-axis, is reduced to $2\pi R_2/s$.  Equivalently, it wraps $r$ times in 
$R_1$-direction; hence the geometric parameter, if measured on the $x_1$-axis 
is reduced to $2\pi R_1/ r$. Thus the position parameter is to be taken modulo 
${2\pi R_1}/r \eq {2\pi R_2}/s$.

This should again be compared to the analysis of \cite{fgrS}. In the case at
hand two parameters have been found: the first couples to the sum of left- and
right-moving momenta, which by the compatibility of the two radii
is required to be quantized in units of $r/{R_1}$. This
nicely fits the position parameter found above. Similarly, there is
a parameter coupling to winding, i.e.\ to the difference of left- and 
right-moving momenta. The latter is quantized in units of $s R_1$, fitting
the quantization of the Wilson lines derived above.

Again one can generalize the analysis to bi-branes that preserve the chiral 
currents only up to automorphisms. If the non-trivial automorphism is taken for
both chiralities, one expects off-diagonal bi-branes; the discussion of the 
parameters largely parallels the one in the preceding paragraphs. In the case 
of different automorphisms, one expects bi-branes filling $S^1_{R_1}\Times 
S^1_{R_2}$, provided that the area of the product space is rational in 
suitable units.
For the specific case $R_2\eq2/{R_1}$ these bi-branes should be related to
defects which implement T-duality. In this context, the fact \cite{hori3}
that the curvature $\pm\, \rmd \theta_1\wedge \rmd \theta_2$ is of the same 
form as the curvature of the Poincar\'e line bundle is highly intriguing.
A careful discussion of this relationship is, again, beyond the scope
of the present paper.


\subsection{WZW bi-branes}

We now turn our attention to bi-branes of WZW models on simply connected
compact Lie groups. Here several new phenomena arise: the
position of possible branes and bi-branes is quantized, and multiplicities
other than zero or one are expected from the algebraic approach.
In fact, from that approach it is known that for these theories
the multiplicities appearing in the fusion of bi-branes as well as the mixed 
fusion of bi-branes and branes are the same as the chiral
fusion multiplicities which are given by the Verlinde formula.

To analyze this issue, it turns out to be convenient to work with fusion
coefficients of type $\caln_{\alpha\beta\gamma}$; here $\alpha$ and
$\gamma$ are group elements characterizing conjugacy classes $\calc_\alpha$ 
and $\calc_\gamma$ of $G$, respectively, which support a brane, while $\beta$ 
is a group element characterizing a bi-brane $\tilde\mu^{-1}(\calc_\beta)$
with $\tilde\mu$ as in \erf{mudef}. In the sequel we assume that all group 
elements are regular, i.e.\ contained in just a single maximal torus of $G$.
We are thus lead to consider the subset 
  \be \bearll
  M_{\alpha\beta\gamma} \!\!\! &
  := p_1^{-1}(\calc_\alpha) \cap \tilde\mu^{-1}(\calc_\beta)
  \cap p_2^{-1}(\calc_\gamma)
  \\{}\\[-.7em] & \,
  = \{ (g_1^{},g_2^{})\iN G\Times G \,|\, g_1^{}\iN\calc_\alpha,\,
  g_2^{}\iN\calc_\gamma,\, g_1^{}g_2^{-1}{\in}\,\calc_\beta \}
  \eear \ee
of $G\Times G$. This set is equipped with a natural $G$-action, obtained by 
combining the adjoint action on $g_1$ and on 
$g_2$. Both branes and bi-branes are equipped with two-forms; as a consequence, 
$M_{\alpha\beta\gamma}$ comes with a natural two-form, namely the sum
  \be
  \omega_{\alpha\beta\gamma}
  := p_1^*\omega_\alpha|^{}_{M_{\alpha\beta\gamma}}
   + p_2^*\omega_\gamma|^{}_{M_{\alpha\beta\gamma}}
   + \bbf_\beta|^{}_{M_{\alpha\beta\gamma}}
  \labl{omegaabc}
of the restrictions of the three two-forms $p_1^*\omega_\alpha$, 
$p_2^*\omega_\gamma$ and $\bbf_\beta$.

According to the results obtained in the algebraic approach, this space should 
be linked to the fusion rules of the chiral WZW theory at level $k$. To see 
how such a relation can exist, we recall that fusion rules are dimensions of 
spaces of conformal blocks. The latter can be obtained by geometric 
quantization from suitable moduli spaces of flat connections; as such they 
arise in the quantization of Chern\hy Simons theories.

The situation relevant for Verlinde multiplicities is given by the 
three-punctured sphere $S^2_{(3)}$, also known as the `pair of pants' 
or trinion. In classical Chern\hy Simons
theory one considers the moduli space of flat connections on $S^2$ whose
monodromy around the three insertion points takes values in conjugacy classes 
$\calc_\alpha$, $\calc_\beta$ and $\calc_\gamma$, respectively. Taking the 
monodromies $g_\alpha\iN \calc_\alpha$, $g_\beta\iN\calc_\beta$ and
$g_\gamma\iN \calc_\gamma$ along circles of the same orientation around all
three insertions, the relations in the fundamental group of the trinion impose
that $g_\alpha g_\beta^{} g_\gamma \eq 1$. Since monodromies are defined only
up to simultaneous conjugation, the moduli space that matters in classical 
Chern\hy Simons theory is isomorphic to the quotient $M_{\alpha\beta\gamma}/G$.

Note that the bounds on the range of bi-branes that appear in the fusion are 
already present before geometric quantization. Indeed, the relevant product 
  \be
  \calc_h * \calc_{h'} := \{ gg' \,|\, g\iN\calc_h,\,g'\iN\calc_{h'} \}
  \labl{class.fus}
of conjugacy classes has already been considered, for $G \eq SU(2)$, in 
\cite{jewe}. It is convenient to characterize a conjugacy class of $SU(2)$ 
by its trace or, equivalently, by the angle $\theta$ with 
  \be
  \cos\theta = \Frac12\, \mathrm{tr}(g) \,,
  \ee
which takes values $\theta\iN[0,\pi]$.
One finds (see Proposition 3.1 of \cite{jewe}) that the (classical) product 
\erf{class.fus} of the two conjugacy classes with angles $\theta,\theta'$ 
is the union of all conjugacy classes with angle $\theta''$ in the range
  \be
  |\theta\,{-}\,\theta'| \le \theta''
  \le \min\{\theta{+}\theta',2\pi\,{-}\,(\theta{+}\theta')\} \,.
  \ee
This already yields the correct upper and lower bounds that appear in the
$SU(2)$ fusion rules.

A full understanding of fusion can only be expected after applying geometric
quantization to the so obtained moduli space: this space must be endowed with
a two-form, which is interpreted as the curvature of a line bundle, and the
holomorphic sections of this bundle are what results from geometric 
quantization. In view of this need for quantization it is a highly non-trivial 
observation that the two-form \erf{omegaabc} furnished by the two branes and 
the bi-brane is exactly the same as the one which arises\,%
  \footnote{~We are grateful to Anton Alekseev for information about this 
  two-form.}
from classical Chern\hy Simons theory.


\section{Outlook} \label{sec5}

Our findings naturally admit various extensions and generalizations.
For instance, one can impose conservation of the currents only up to an 
automorphism of the horizontal Lie algebra, which may be chosen independently 
for left- and right-moving degrees of freedom. Also, our methods can be 
clearly extended to more general classes of conformal field theories,
in particular to WZW models on non-simply connected groups, coset models, 
as well as to theories of several free bosons compactified on a torus and 
to orbifolds thereof, including asymmetric orbifolds such as lens spaces. 
Another generalization concerns defects which separate sigma models on 
two different Lie groups that share the same Lie algebra.

Furthermore, our results provide independent evidence for the idea that 
there is an intimate relation between defects and correspondences. This idea 
has played a role in a field theoretic realization of the geometric Langlands 
program (see Section 6.4 of \cite{kawi}). 
It is therefore not unreasonable to expect that defects and, more generally, 
the algebraic and categorical structure behind RCFT correlators,
will enter in a CFT-inspired approach to the Langlands program.

Finally it could be rewarding to unravel similar structures in lattice models.

\vskip4em

\noindent
{\bf Acknowledgements.}\\
We thank Birgit Richter for a helpful discussion and Anton Alekseev for a
helpful correspondence. J.F.\ is partially supported by VR under project no.\ 
621-2006-3343. C.S.\ and K.W.\ are supported by the Sonderforschungsbereich
``Particles, Strings and the Early Universe - the Structure of Matter and 
Space-Time''.

\vskip2em
\newpage
\appendix

\section{Birelative (co-)homology} \label{appA}

In this Appendix we discuss the well-definedness of the Wess\hy Zumino term 
\erf{neu1} in the presence of a defect line. To this end we set up a homology 
theory based on singular homology, which can be understood as a generalization 
of relative homology, and which we will accordingly call birelative homology.
The associated cohomology theory with real coefficients can be identified 
with a cohomology theory based on differential forms, which we call birelative 
de Rham cohomology. These structures enable us to formulate precise conditions
under which the Wess\hy Zumino term \erf{neu1} is well-defined up to integers.

Recall that the (singular) homology $H_k(M)$ of a smooth manifold $M$ is the 
homology of the singular chain complex with chain groups $\Delta_k(M)$, 
consisting of (smooth) $k$-simplices in $M$ and boundary operator 
$\partial{:}\ \Delta_k(M) \,{\to}\, \Delta_{k-1}(M)$ (we suppress the index 
of the boundary operator $\partial$, as it can be inferred from the index of 
the simplex on which it acts). If $Q \,{\subset}\, M_1 \Times M_2$ is a
submanifold, we define the $k$th {\em birelative chain group\/} 
of the triple $(M_1,M_2,Q)$ to be
  \be
  \Delta_k(M_1,M_2,Q):= \Delta_k(M_1)
  \oplus \Delta_k(M_2) \oplus \Delta_{k-1}(Q) \,\text{.}
  \ee
Using the projections $p_i{:}\ M_1 \Times M_2 \,{\to}\, M_i$ and the inclusion
map $\iota{:}\ Q \,{\hookrightarrow}\, M_1 \Times M_2$, and the induced
chain maps $(p_i)_{*}$ and $\iota_{*}$, we define the homomorphism
  \be
  \begin{array}{lrcl}
  \partial:&\Delta_k(M_1,M_2,Q)&\!\!\to\!\!&\Delta_{k-1}(M_1,M_2,Q)
  \\[2pt]
  &(\sigma_1,\sigma_2,\tau)&\!\!\mapsto\!\!&
  (\partial\sigma_1\,{+}\,(p_1)_{*}\iota_{*}\tau,
  \partial\sigma_2\,{-}\,(p_2)_{*}\iota_{*}\tau,-\partial\tau) \,\text{.}
  \end{array}
  \ee
It is easy to verify that this map satisfies $\partial^2 \eq 0$, i.e.\ we have
endowed the birelative chain groups with the structure of a
complex. We call its homology groups the {\em birelative homology groups\/} 
and denote them by $H_k(M_1,M_2,Q)$. Explicitly, an element of $H_k(M_1,M_2,Q)$ 
is represented by a triple $(\sigma_1,\sigma_2,\tau)$ of chains $\sigma_i \iN 
\Delta_k(M_i)$, $i\eq1,2$, and a cycle $\tau\iN\Delta_{k-1}(Q)$, such that 
$\partial\sigma_1 \eq (p_1)_{*}\iota_{*}\tau$ and $\partial\sigma_2\eq 
{-}(p_2)_{*}\iota_{*}\tau$. For each degree $k$, the birelative chain 
group fits, by definition, into the short exact sequence
  \be
  \alxydim{@C=0.8cm}{0 \ar[r] & \Delta_k(M_1)\,{\oplus}\,\Delta_k(M_2) \ar[r]^-{\alpha} 
  & \Delta_k(M_1,M_2,Q) \ar[r]^-{\beta} & \Delta_{k-1}(Q) \ar[r]& 0 \,,}
  \labl{neu6}
in which $\alpha$ is the inclusion and $\beta$ is the projection.
These induce a long exact sequence 
  \be
  \alxydim{@C=0.38cm}{... \ar[r]^-{} & H_k(M_1) \,{\oplus}\, H_k(M_2)
  \ar[r]^-{}
  & H_k(M_1,M_2,Q) \ar[r]^{} & H_{k-1}(Q) \ar[r] & H_{k-1}(M_1) \,{\oplus}\,
  H_{k-1}(M_2) \ar[r] &...}
  \labl{neu8}
in homology. 

\medskip

To explain the term birelative homology we observe that we have
generalized relative homology in the following sense: if we
take $M_2\eq pt$, so that we can identify $Q$ with a submanifold of $M_1$, 
then there is a canonical isomorphism $H_{k}(M_1,pt,Q) \,{\to}\, H_{k}(M_1,Q)$.
Here $H_{k}(M_1,Q)$, the relative homology group of $M_1$ with respect to the 
submanifold $Q$, is constructed as the homomorphism 
$[(\sigma_1,\sigma_2,\tau)]\,{\mapsto}\,[\sigma_1]$ which can be shown to be an 
isomorphism by using the 5-lemma (see e.g.\ \cite{bredon}, Lemma IV.5.10) 
applied to the exact sequence \erf{neu8} and the corresponding sequence
in relative homology.  

\medskip

Dual to the singular homology groups there are singular cohomology groups,
defined to be the cohomology of a complex whose cochain groups are
  \be
  \Delta^k(M,R) := \mathrm{Hom}(\Delta_k(M),R)
  \ee
for a coefficient ring $R$, and whose coboundary operator 
  \be
  \delta:\quad \Delta^{k}(M,R) \to \Delta^{k+1}(M,R)
  \end{equation}
is given by $\delta\varphi(\sigma) \,{:=}\, \varphi(\partial \sigma)$ for
any $(k{+}1)$-simplex $\sigma$ in $M$. There is a canonical pairing
  \be
  \label{neu3}
  H^k(M,R) \Times H_k(M) \to R ~~~\text{ with }~~~
  ([\varphi],[\sigma]) \mapsto \varphi(\sigma) \,\text{,}
  \ee
which is easily seen to be well defined. It is often convenient to recover the
cohomology groups with values in the real numbers in a geometric way, for instance
through differential forms. Let us recall how this works. The integrals of 
$k$-forms $\varphi\iN\Omega^k(M)$ over $k$-simplices $\sigma\iN \Delta_k(M)$ 
define homomorphisms $\Psi_k{:}\ \Omega^k(M) \,{\to}\, \Delta^k(M,\R)$ which, 
by Stokes' theorem, fit together to a chain map. The induced homomorphism
  \be
  \Psi^{*}:\quad H^k_{\mathrm{dR}}(M) \to H^k(M,\R)
  \labl{neu4}
from de Rham cohomology to singular cohomology is an isomorphism, which is known
as the de Rham isomorphism (see e.g.\ Theorem V.9.1 of \cite{bredon}). 

\medskip

Analogously as for ordinary singular cohomology, we can also define birelative 
cohomology. Thus there are birelative cochain groups $\Delta^k(M_1,M_2,Q,R)$,
birelative cohomology groups $H^k(M_1,M_2,Q,R)$, and a canonical pairing
  \be
  H^k(M_1,M_2,Q,R) \times H_k(M_1,M_2,Q) \to R \,\text{.}
  \labl{neu5}
Note that because the exact sequence \erf{neu6} splits, the dual sequence
  \be
  \alxydim{@C=0.8cm}{0 \ar[r] & \Delta^{k-1}(Q,R)
  \ar[r] & \Delta^k(M_1,M_2,R) \ar[r]
  & \Delta^k(M_1) \,{\oplus}\, \Delta^k(M_2) \ar[r] & 0}
  \labl{neu9}
is exact, too, and induces a long exact
sequence in cohomology. We would like be able to express the birelative 
cohomology groups with real coefficients by
differential forms in a similar way as the de Rham isomorphism does it for ordinary
cohomology. To this end we consider the vector spaces
  \be
  \Omega^k(M_1,M_2,Q) := \Omega^k(M_1) \oplus \Omega^k(M_2) \oplus
  \Omega^{k-1}(Q)
  \ee
together with the linear maps
  \be
  \begin{array}{lrcl}
  \mathrm{d}:&\Omega^k(M_1,M_2,Q)&\!\!\to\!\!&\Omega^{k+1}(M_1,M_2,Q)
  \\{}\\[-.8em]
  &(H_1,H_2,\bbf)&\!\!\mapsto\!\!&
  (\mathrm{d}H_1^{},\mathrm{d}H_2^{},\iota^{*}(p_1^{*}H_1^{}{-}p_2^{*}H_2^{})
  \,{-}\, \mathrm{d}\bbf) \,\text{.}
  \eear \ee
This indeed defines a complex:
  \be \bearll
  \mathrm{d}^2(H_1^{},H_2^{},\bbf) \!\!
  &= \mathrm{d}\,(\mathrm{d}H_1^{},\,\mathrm{d}H_2^{},\,\iota^{*}
  (p_1^{*}H_1^{} - p_2^{*}H_2^{}) - \mathrm{d}\bbf)
  \\{}\\[-.7em]
  & = (\mathrm{d}^2H_1^{},\,\mathrm{d}^2H_2^{},\,
  \iota^{*}(p_1^{*}\mathrm{d}H_1^{} - p_2^{*}\mathrm{d}H_2^{})
  - \mathrm{d}\iota^{*}(p_1^{*}H_1^{} - p_2^{*}H_2^{}) + \mathrm{d}^2\bbf)
  \\{}\\[-.7em]
  & = (0,0,0) \,.
  \eear \ee
We call the cohomology of this complex the \textit{birelative de Rham cohomology\/} 
and denote it by $H^k_\text{dR}(M_1,M_2,Q)$. By putting
$M_2 \,{=}\, pt$, this is nothing but the relative de Rham cohomology 
of the map $\iota{:}\ Q \,{\to}\, M$, see e.g.\ I $\S$6 of \cite{BOtu}. 

Notice that a simply connected $M_1$-$M_2$-bi-brane $(Q,\bbf)$ provides us 
with an element $(H_1,H_2,\bbf)$ of $\Omega^3(M_1,M_2,Q)$. The condition 
\erf{4.1} on the two-form $\bbf$ on the bi-brane shows that $(H_1,H_2,\bbf)$ 
is closed and thus defines a class in the birelative de Rham cohomology. 

\medskip

Similarly to the definition of the homomorphism $\Psi{:}\ \Omega^k(M) \,{\to}\,
\Delta^k(M,\R)$ mentioned above we obtain a natural homomorphism
  \be
  \Psi_{\mathrm{bi}}:\quad \Omega^k(M_1,M_2,Q)
  \to \Delta^k(M_1,M_2,Q,\R)
  \ee
which by definition associates to a triple $(H_1,H_2,\bbf)\iN\Omega^k(M_1,M_2,Q)$
evaluated on an element $(\sigma_1,\sigma_2,\tau)\iN\Delta_k(M_1,M_2,Q)$
the real number
  \be
  \Psi_{\mathrm{bi}}(H_1,H_2,\bbf)(\sigma_1,\sigma_2,\tau)
  := \int_{\!\sigma_1}H_1 + \int_{\!\sigma_2}H_2 + \int_{\!\tau}\bbf\,\text{.} 
  \labl{neu100}
The homomorphisms $\Psi_{\mathrm{bi}}$ fit together to a chain map: 
  \begin{eqnarray}
  & (\mathrm{\delta}\Psi_{\mathrm{bi}}(H_1,H_2,\bbf))(\sigma_1,\sigma_2,\tau)
  \!\!
  &= \Psi_{\mathrm{bi}}(H_1,H_2,\bbf)(\partial\sigma_1
  {+}(p_1)_{*}\iota_{*}\tau,\partial\sigma_2{-}(p_2)_{*}\iota_{*}\tau,-\partial\tau)
  \nonumber\\
  &&=\displaystyle\int_{\!\partial\sigma_1 \,+ (p_1)_{*}\iota_{*}\tau}\! H_1
  \,+ \int_{\!\partial\sigma_2-(p_2)_{*}\iota_{*}\tau}\! H_2
  \,+ \int_{\!-\partial\tau}\! \bbf
  \nonumber\\
  &&=\int_{\!\sigma_1}\!\mathrm{d}H_1 \,+ \int_{\!\sigma_2}\!\mathrm{d}H_2
  \,+ \int_{\!\tau}\iota^{*}(p_1^{*}H_1^{}{-}p_2^{*}H_2^{}) \,{-}\, \mathrm{d}\bbf
  \nonumber\\[3pt]
  &&= \Psi_{\mathrm{bi}}(\mathrm{d}H_1,\mathrm{d}H_2,\iota^{*}(p_1^{*}H_1^{}
  {-} p_2^{*}H_2^{}) \,{-}\, \mathrm{d}\bbf)(\sigma_1,\sigma_2,\tau)
  \nonumber\\[4pt]
  &&= \Psi_{\mathrm{bi}}(\mathrm{d}(H_1,H_2,\bbf))(\sigma_1,\sigma_2,\tau)
  \,\text{.}
  \end{eqnarray}
We infer that the induced homomorphism
  \be
  \Psi^{*}_{\mathrm{bi}}:\quad
  H^{k}_{\mathrm{dR}}(M_1,M_2,Q) \to H^k(M_1,M_2,Q,\R) 
  \labl{neu10}
is an isomorphism, analogously as the de Rham isomorphism. To prove this claim, 
note that by definition we have an exact sequence
  \be
  \alxydim{@C=0.8cm}{0 \ar[r] & \Omega^{k-1}(Q) \ar[r]^-{\alpha}
  & \Omega^k(M_1,M_2,Q) \ar[r]^-{\beta}
  & \Omega^k(M_1) \,{\oplus}\, \Omega^k(M_2) \ar[r] & 0\,\text{,}}
  \ee
where $\alpha(\bbf) \,{:=}\, (0,0,\bbf)$ and $\beta(H_1,H_2,\bbf)
\,{:=}\, (H_1,H_2)$. It induces a long exact sequence 
  \be
  \alxydim{@C=0.67cm}{... \ar[r] & H^{k-1}_{\mathrm{dR}}(Q)
  \ar[r]^-{\alpha^{*}} &H^{k}_{\mathrm{dR}}(M_1,M_2,Q)
  \ar[r]^-{\beta^{*}} & H^{k}_{\mathrm{dR}}(M_1)
  \,{\oplus}\, H^{k}_{\mathrm{dR}}(M_2) \ar[r]^-{\delta}
  & H^{k}_{\mathrm{dR}}(Q) \ar[r]& ...}
  \ee
in (birelative) de Rham cohomology. Together with the long exact sequence
in birelative cohomology with values in $\R$, induced by the exact
sequence (\ref{neu9}), we have the following diagram with exact rows:
  \begin{equation*}
  \alxydim{@C=0.38cm}{
  H^{k-1}_{\mathrm{dR}}(M_1) {\oplus} H^{k-1}_{\mathrm{dR}}(M_2)
    \ar[d]_{\Psi^{*} \oplus \Psi^{*}} \ar[r]
  & H^{k-1}_{\mathrm{dR}}(Q) \ar[d]_{\Psi^{*}}\ar[r]
  & H^{k}_{\mathrm{dR}}(M_1,M_2,Q) \ar[d]|{\Psi^{*}_{\mathrm{bi}}} \ar[r]
  & H^{k}_{\mathrm{dR}}(M_1) {\oplus} H^{k}_{\mathrm{dR}}(M_2)
    \ar[d]^{\Psi^{*} \oplus \Psi^{*}} \ar[r]
  & H^{k}_{\mathrm{dR}}(Q) \ar[d]^{\Psi^{*}}
  \\
  \txt{$H^{k-1}(M_1,\R)\qquad\quad$\\$\qquad\oplus H^{k-1}(M_2,\R)$} \ar[r]
  & H^{k-1}(Q,\R)\ar[r] &H^{k}(M_1,M_2,Q,\R) \ar[r]
  & \txt{$H^{k}(M_1,\R)\qquad$\\$\qquad\oplus H^{k}(M_2,\R)$} \ar[r]
  & H^{k}(Q,\R) }
  \end{equation*}
It is easy to check that all subdiagrams commute, so that the 5-lemma 
implies that $\Psi_{\mathrm{bi}}^{*}$ is an isomorphism.  

\medskip

In the same way as for ordinary cohomology, we say that a cocycle in
$\Omega^k(M_1,M_2,Q)$ is integral iff its class -- identified
by $\Psi^{*}_{\mathrm{bi}}$ with a class in $H^k(M_1,M_2,Q,\R)$
-- lies in the image of the induced homomorphism
  \be
  H^k(M_1,M_2,Q,\Z) \to H^k(M_1,M_2,Q,\R) \,\text{.}
  \ee
In this case the canonical pairing \erf{neu5} of 
$\Psi_{\mathrm{bi}}^{*}([H_1,H_2,\bbf])$ with  any birelative
homology class $[(\sigma_1,\sigma_2,\tau)]$, which is given by
  \be
  \int_{\!\sigma_1}H_1 + \int_{\!\sigma_2}H_2
  + \int_{\!\tau}\bbf\,\text{,}
  \ee
is an integer. Analogously as for WZW models in the bulk and on the boundary 
of a world sheet, this notion of integral classes is essential to achieve the 
well-definedness of Wess\hy Zumino terms. We infer the following result:

\def\leftmargini{\parindent}\begin{itemize}
\item[{}]
The Wess\hy Zumino term $S[\phi_1,\phi_2]$ \erf{neu1} of a simply connected 
$M_1$-$M_2$-bi-brane $(Q,\bbf)$ is well-defined up to integers,
provided that the class of $(H_1,H_2,\bbf)$ in the birelative
de Rham cohomology group $H^3_\text{dR}(M_1,M_2,Q)$ is integral.
\end{itemize}

To prove this claim, recall that the definition of  $S[\phi_1,\phi_2]$ involves
choices of submanifolds $D$ of $Q$ and $B_i$ of $M_i$. If we represent
these submanifolds as singular chains, then
  \be
  \partial D = \phi_S(S)
  \,\text{, }\quad~
  \partial B_1 = \phi_1(\Sigma_1) - (p_1)_{*}D
  \quad~\text{ and }\quad
  \partial B_2 = \phi_2(\Sigma_2) + (p_2)_{*}D \,\text{.}
  \ee
Consider now different choices $D'$, $B_1'$ and $B_2'$, and let
$\tau \,{:=}\, D\,{-}\,D'$ be a chain in $\Delta_2(Q)$ and 
$\sigma_i \,{:=}\, B_i\,{-}\,B_i'$ be chains in $\Delta_3(M_i)$. We find
  \be
  \partial\tau = 0 \,\text{, }\quad~
  \partial\sigma_1=-(p_1)_{*}\tau \quad~\text{ and } \quad
  \partial\sigma_2=  (p_2)_{*}\tau \,,
  \ee
so that $(\sigma_1,\sigma_2,\tau)$ is a cycle in the birelative homology
$H_3(M_1,M_2,Q)$. The ambiguities of the Wess\hy Zumino term
$S[\phi_1,\phi_2]$ are thus of the form
  \be
  \Big( \int_{\!B_1}\! H_1^{} 
  + \!\int_{\!B_2} \!H_2^{} + \!\int_{\!D}\! \bbf \Big)
  - \Big( \!\int_{\!B'_1}\! H_1^{}- \!\int_{\!B'_2}\! H_2^{}
    + \!\int_{\!D'}\! \bbf \Big)
  = \int_{\sigma_1}\! H_1^{} + \int_{\sigma_2}\! H_2^{} + \int_{\tau}\!\bbf
  \,\text{.}
  \labl{neu7}
In view of \erf{neu100} the ambiguities (\ref{neu7}) are nothing but the 
pairing of the cycle $(\sigma_1,\sigma_2,\tau)$ with $(H_1,H_2,\bbf)$. If
$(H_1,H_2,\bbf)$ is integral, this gives an integer.


\section{Bundle gerbes and defects} \label{sec:BG} \label{appB}

As we have explained
in section \ref{secWZ} it is perfectly accurate to characterize
bundle gerbes on 2-connected target spaces $M_1$ and $M_2$ by their
curvature three-forms $H_1$ and $H_2$. Under this condition, we have defined an
$M_1$-$M_2$-bi-brane to be a simply connected submanifold
$Q$ of $M_1 \Times M_2$ together with a two-form $\bbf$ on $Q$ that obeys
  \be
  p_1^{*}H|_Q^{} = p_2^{*}H|_Q^{} + \mathrm{d}\bbf \,\text{.}
  \labl{2neu3}
In this Appendix we generalize this definition to bi-branes between target
spaces with are not 2-connected. This makes it necessary to work with the full
structure of a hermitian bundle gerbe. Examples of non-2-connected target 
spaces are provided by non-simply connected Lie groups, such as the group
$SO(4n)/\Z_2$, which admits two non-isomorphic bundle gerbes with
the same curvature three-form $H$. At the same time, we drop the
restriction on the bi-brane $Q$ to be simply connected. Examples
of non-simply connected bi-branes are provided by certain
biconjugacy classes of non-simply connected Lie groups.

\subsection{Gerbe modules}

Let us first recall how branes have been understood using bundle gerbes
\cite{gawedzki4}. Let $\mathcal{G}$ be a bundle gerbe on the target space
$M$ with curvature $H$.  The geometric structure related to a conformal 
boundary condition consists of a pair\,%
  \footnote{~But not every such pair corresponds to a conformal boundary 
  condition; there are far more such pairs than conformal boundary conditions.} 
$(Q,\mathcal{E})$, with $Q$ a submanifold of $M$ and $\mathcal{E}$ 
a gerbe module for the restriction of $\mathcal{G}$ to $Q$.
Such gerbe modules are vector bundles twisted by
the bundle gerbe $\mathcal{G}$. We can view them as bundle gerbe morphisms
  \be
  \mathcal{E}:\quad \mathcal{G}|_Q^{} \to \mathcal{I}_{\omega}
  \ee
from $\mathcal{G}|_Q$ to a trivial bundle gerbe $\mathcal{I}_{\omega}$ given
by a two-form $\omega$ on $Q$ \cite{waldorf1}. The two-form $\omega$ is called
the curvature of the gerbe module. A necessary condition for the existence
of the morphism $\mathcal{E}$ is the equality
  \be
  H|_Q^{} = \mathrm{d}\omega
  \labl{2neu1}
on $Q$. If the submanifold $Q$ is not simply connected, then non-trivial flat 
line bundles exist. Since gerbe modules (of equal rank) with the same curvature 
$\omega$ form a torsor over the group of flat line bundles, in this situation 
non-isomorphic gerbe modules with the same curvature exist.
This happens, for example, for the equatorial conjugacy class of $SO(3)$, which
has the topology of $\RP^2$ and thus admits two non-isomorphic flat line
bundles, whose action relates two non-isomorphic gerbe modules.

\medskip

The arguably most direct way to understand (hermitian) bundle gerbes (with
connective structure) is in terms of their local data: with respect to a 
good open cover $\mathfrak{U} \eq\lbrace U_i \rbrace_{i\in I}$ of $M$, 
a bundle gerbe $\mathcal{G}$
can be described by a collection $(g_{ijk},A_{ij},B_i)$ of smooth functions
$g_{ijk}{:}\ U_i \,{\cap}\, U_j \,{\cap}\, U_k \,{\to}\, U(1)$, 1-forms 
$A_{ij}\iN \Omega^1(U_i \,{\cap}\, U_j)$ and two-forms $B_i \iN \Omega^2(U_i)$, 
satisfying the cocycle conditions
  \be \begin{array}{rcl}
  g_{jkl}^{-1} \cdot g_{ikl}^{} \cdot g_{ijl}^{-1}\cdot g_{ijk}^{} &\!\!=\!\!&
  1 \quad\text{ on }~ U_i \cap U_j \cap U_k \cap U_l \,,
  \\{}\\[-.7em]
  -\mathrm{i}\,g_{ijk}^{-1}\rmd g_{ijk}^{}+A_{jk}^{}-A_{ik}^{}+A_{ij}^{}
  &\!\!=\!\!& 0\quad\text{ on }~U_i \cap U_j \cap U_k \,,
  \\{}\\[-.7em]
  \mathrm{d}A_{ij}^{}-B_{j}^{} + B_i^{}
  &\!\!=\!\!& 0\quad\text{ on }~U_i \cap U_j \,\text{.}
  \eear \ee
The curvature of $\mathcal{G}$ is the globally defined three-form $H$ with
$H|_{U_i} \,{:=}\, \mathrm{d}B_i$. For example, the local data of the trivial 
bundle gerbe $\mathcal{I}_{\omega}$ are $(1,0,\omega|_{U_i \cap Q})$.
A rank-$n$ bundle gerbe module $\mathcal{E}{:}\ \mathcal{G}|_Q \,{\to}\,
\mathcal{I}_{\omega}$ is in this formalism described by a collection
$(G_{ij},\Pi_i)$ of smooth functions $G_{ij}{:}\ U_i \,{\cap}\, U_j \,{\cap}\, 
Q \,{\to}\, U(n)$ and $\mathfrak{u}(n)$-valued 1-forms 
$\Pi_i \iN \Omega^1(U_i\cap Q) \oti \mathfrak{u}(n)$ which relate the local 
data of the bundle gerbes $\mathcal{G}|_Q$ and $\mathcal{I}_{\omega}$
in the following way:
  \be
  \begin{array}{lcll}
  1 &\!\!=\!\!& g_{ijk}^{} \,\cdot\, G_{ik}^{}\,G_{jk}^{-1}\,G_{ij}^{-1}
  &~\text{ on }~Q \cap U_i \cap U_j \cap U_k \,,
  \\{}\\[-.7em]
  0 &\!\!=\!\!& A_{ij}^{} + \Pi_{j}^{} - G_{ij}^{-1}\Pi_{i}^{}\,G_{ij}^{} -
  \mathrm{i}\,G_{ij}^{-1}\,\mathrm{d}G_{ij}^{}
  &~\text{ on }~Q \cap U_i \cap U_j \,,
  \\{}\\[-.9em]
  \omega &\!\!=\!\!& \,B_i^{}\, + \mbox{$\displaystyle \Frac{1}{n}$}\, 
  \mathrm{tr}(\mathrm{d}\Pi_i^{})&~\text{ on }~Q \cap U_i \,\text{.}
  \label{2neu2}
  \eear \ee
Note that the derivative of the last equality reproduces the relation
(\ref{2neu1}). Also note that if the bundle gerbe $\mathcal{G}$ is itself
trivial, i.e.\ has local data $(1,0,B|_{U_i})$ for a globally defined
Kalb\hy Ra\-mond field $B \iN \Omega^2(M)$, then $(G_{ij},\Pi_{i})$ are the
local data of a rank-$n$ vector bundle over $Q$ with curvature of trace 
$n\,(\omega{-}B)$. This explains the terminology
\textquotedblleft twisted\textquotedblright\ vector bundle in the non-trivial 
case. Finally, notice that if one changes $(G_{ij},\Pi_{i})$ by local data of a
non-trivializable flat vector bundle over the world volume $Q$ of the bi-brane,
then one obtains a new bundle gerbe module with the same
curvature. In this way the existence of non-trivial flat vector bundles 
over $Q$ makes the use of bundle gerbe modules unavoidable.

\medskip

In the case of WZW conformal field theories with $M\eq G$ one considers
in particular so-called symmetric branes, which preserve the current algebra 
in the presence of boundaries, and thus in particular conformal invariance.
Symmetric D-Branes $(Q,\mathcal{E})$ can be characterized by three conditions 
\cite{gawedzki4}: 
\\[-1.7em]
\def\leftmargini{1.3em}\begin{enumerate} \addtolength\itemsep{-5pt}
\item 
the world volume $Q$ of the brane is a conjugacy class $\mathcal C_h$ of $G$;
\item
the local two-forms $\mathrm{d}\Pi_i$ take their values only in the center of
the Lie algebra $\mathfrak{u}(n)$ and can thus be identified with real two-forms;
\item
the two-form $\omega$ is fixed to
  \be
  \omega = \Big \langle \theta|_{\mathcal{C}_h}^{} \wedge \frac{\mathrm{Ad}^{-1}
  + 1}{\mathrm{Ad}^{-1}-1}\,\theta|_{\mathcal{C}_h}^{} \Big \rangle\,.
  \ee
\end{enumerate}
The conditions 2 and 3 restrict the choice of the conjugacy class to conjugacy 
classes that correspond to integrable highest weights. This amounts in 
particular to having a finite number of non-intersecting brane world volumes.


\subsection{Gerbe bimodules}

That bundle gerbe modules are the appropriate structure for branes in the case
of non-2-connected target spaces or non-simply connected supports, 
together with the folding trick suggests the corresponding structure as the 
appropriate generalization for bi-branes: for bundle gerbes $\mathcal{G}_1$ 
and $\mathcal{G}_2$  over $M_1$ and $M_2$, an {\em $M_1$-$M_2$-bi-brane\/} 
is a submanifold $Q \,{\subset}\, M_1 \Times M_2$ together with a 
$(p_1^{*}\mathcal{G}_1)|_Q$-$(p_2^{*}\mathcal{G}_2)|_Q$-bimodule:
a bundle gerbe morphism
  \be
  \mathcal{D} :\quad
  (p_1^{*}\mathcal{G}_1^{})|_Q^{} \to (p_2^{*}\mathcal{G}_2^{})|_Q^{}
  \otimes \mathcal{I}_{\bbf}
  \ee
with $\bbf$ as in \erf{2neu3}. Here we shall call the two-form $\bbf$ the 
curvature of the bimodule. This definition is related to the folding trick in 
the sense, that -- using the appropriate notion of duality for bundle gerbes
(see section 1.4 of \cite{waldorf1}) -- a $\mathcal{G}_1$-$\mathcal{G}_2$-bimodule
is the same as a $(\mathcal{G}_1{\otimes}\mathcal{G}_2^{*})$-module. 

\medskip

To consider a bundle gerbe bimodule $\mathcal{D}$ in
the local data formalism, let $\mathfrak{U}$ be a good covering of $M_1 \Times
M_2$, let $(g_{ijk},A_{ij},B_i)$ be local data of $p_1^{*}\mathcal{G}_1^{}$,
and $(g_{ijk}',A_{ij}',B_i')$ local data of $p_2^{*}\mathcal{G}_2^{}$.
Then the bimodule has local data $(G_{ij},\Pi_i)$ similar to a bundle gerbe
module, but now satisfying
  \be
  \begin{array}{rcll}
  g_{ijk}' &\!\!=\!\!& g_{ijk}^{} \,\cdot\, G_{ik}^{}\,G_{jk}^{-1}\,G_{ij}^{-1}
  &~\text{ on }~Q \cap U_i \cap U_j \cap U_k \,,
  \\{}\\[-.7em]
  A_{ij}' &\!\!=\!\!& A_{ij}^{} + \Pi_{j}^{} - G_{ij}^{-1}\Pi_{i}^{}\,G_{ij}^{} -
  \mathrm{i}\,G_{ij}^{-1}\,\mathrm{d}G_{ij}^{}
  &~\text{ on }~Q \cap U_i \cap U_j \,,
  \\{}\\[-.9em]
  B_i'\,{+}\,\bbf &\!\!=\!\!& \,B_i^{}\, + \Frac{1}{n}\, 
  \mathrm{tr}(\mathrm{d}\Pi_i^{})&~\text{ on }~Q \cap U_i \,\text{.}
  \end{array}
  \labl{3neu2}
Again we make three observations: First, the derivative of the third equality 
gives equation (\ref{2neu3}); second, if both bundle gerbes
$p_1^{*}\mathcal{G}_1^{}$ and $p_2^{*}\mathcal{G}_2^{}$ are trivial, then a 
bimodule is just a rank-$n$ vector bundle over $Q$ with curvature of trace
$n\,(B'{-}B{+}\bbf)$; and third, we can still change
the local data $(G_{ij},\Pi_{i})$ by local data of a flat vector bundle over $Q$
and obtain another bimodule with the same curvature. Such phenomena arise, in 
particular, for bi-branes for WZW theories on non-simply connected Lie groups.


\subsection{Holonomy in the presence of defects}

We have generalized the definition of bi-branes from simply connected
bi-branes between 2-connected target spaces with three-forms to arbitrary
bi-branes between arbitrary target spaces with bundle gerbes. Now we shall 
generalize the Wess\hy Zumino term for bi-branes as given in \erf{neu1} 
to the general case as well.

\medskip

Let $M_1$ and $M_2$ be smooth manifolds with bundle gerbes $\mathcal{G}_1$
and $\mathcal{G}_2$ respectively, and let $(Q,\mathcal{E})$ be a bi-brane,
i.e.\ a submanifold $Q$ of $M_1\Times M_2$ together with a
$(p_1^{*}\mathcal{G}_1^{})|_Q$-$(p_2^{*}\mathcal{G}_2^{})|_Q$-bi\-mo\-dule
  \be
  \mathcal{D} :\quad (p_1^{*}\mathcal{G}_1^{})|_Q^{}
  \to (p_2^{*}\mathcal{G}_2^{})|_Q^{} \otimes \mathcal{I}_{\bbf}
  \ee
with curvature $\bbf$.
Recall that we defined  the Wess\hy Zumino term for the following situation:
a closed oriented world sheet $\Sigma$ with an embedded oriented circle 
$S\,{\subset}\, \Sigma$, which separates the world sheet into two
components, $\Sigma \eq \Sigma_1 \,{\cup_{S}}\, \Sigma_2$, together with  maps
$\phi_i{:}\ \Sigma_i \,{\to}\, M_i$ for $i\eq1,2$
such that the image of the combined map
  \be
  \begin{array}{lrcl} 
  \phi_S:& S &\!\!\to\!\!& M_1 \Times M_2 \\
  & s &\!\!\mapsto\!\!& (\phi_1(s) , \phi_2(s))
  \eear \ee
is contained in $Q$. The orientation of $\Sigma_i$ is the one inherited
from the orientation of $\Sigma$, and without loss of generality we take 
$\partial \Sigma_1 \eq S$ and $\partial \Sigma_2 \eq \overline{S}$.

\medskip

To define the Wess\hy Zumino term we use the formalism introduced in
\cite{waldorf1},
which emphasizes the role of morphisms between bundle gerbes, in particular
between trivial bundle gerbes. According to \cite{waldorf1},
equivalence classes of morphisms
$\mathcal{A}{:}\ \mathcal{I}_{\rho_1} \,{\to}\, \mathcal{I}_{\rho_2}$ are in 
natural bijection with equivalence classes of hermitian vector bundles $E$ with 
connection whose curvature satisfies
  \be
  \Frac{1}{n}\,\mathrm{tr}(\mathrm{curv}(E)) = \rho_2 - \rho_1 \,\text{,}
  \ee
with $n$ the rank of $E$. We write $\mathrm{Bun}(\mathcal{A})$ for the
vector bundle corresponding to the morphism $\mathcal{A}$. This assignment
has three important properties (Proposition 4 in \cite{waldorf1}):
\begin{itemize}
\addtolength\itemsep{-7pt}
\item 
if the morphism $\mathcal{A}$ is invertible, then the vector bundle 
$\mathrm{Bun}(\mathcal{A})$ is of rank one, i.e.\ a line bundle; furthermore
  \be
  \mathrm{Bun}(\mathcal{A}^{-1}) = \mathrm{Bun}(\mathcal{A})^{*}\,\text{;}
  \ee
\item
it is compatible with the composition of morphisms,
  \be
  \mathrm{Bun}(\mathcal{A}' \,{\circ}\, \mathcal{A}) = \mathrm{Bun}(\mathcal{A})
  \otimes \mathrm{Bun}(\mathcal{A}')
  \quad\text{ and }\quad
  \mathrm{Bun}(\id_{\mathcal{I}_{\rho}})= 1 \,\text{;}
  \labl{3neu1}
\item
it is compatible with tensor products,
  \be
  \mathrm{Bun}(\mathcal{A}' \oti \mathcal{A}) = \mathrm{Bun}(\mathcal{A})
  \otimes \mathrm{Bun}(\mathcal{A}') \,\text{.}
  \labl{4neu2}
\end{itemize}
As an illustration, consider a manifold $M$ with two bundle gerbes $\mathcal{G}_1$
and $\mathcal{G}_2$, and a $\mathcal{G}_1$-$\mathcal{G}_2$-bi\-mo\-dule
$\mathcal{D}{:}\ \mathcal{G}_1 \,{\to}\, \mathcal{G}_2 \oti \mathcal{I}_{\omega}$.
Suppose we have trivializations of each of the bundle gerbes $\mathcal{G}_1$ and 
$\mathcal{G}_2$, i.e.\ bundle gerbe isomorphisms $\mathcal{T}_i{:}\ \mathcal{G}_i 
\,{\to}\, \mathcal{I}_{\rho_i}$. By composition, we obtain a bundle gerbe morphism
  \be
  \widetilde{\mathcal{D}} := (\mathcal{T}_2^{} \oti
  \id_{\mathcal{I}_{\omega}}) \circ \mathcal{D} \circ \mathcal{T}_1^{-1} :\quad
  \mathcal{I}_{\rho_1} \to \mathcal{I}_{\rho_2 + \omega} \,\text{.}
  \labl{3neu7}
It corresponds to a vector bundle 
$E \,{:=}\, \mathrm{Bun}(\widetilde{\mathcal{D}})$ over $M$. 
Summarizing, a gerbe bimodule together with trivializations gives a hermitian
vector bundle on $M$ with connection. Let us discuss how the vector bundle $E$ 
depends on the choice of the trivializations. If $\mathcal{T}_1'$ and 
$\mathcal{T}_2'$ are two different choices of trivializations and $\widetilde
{D}'$ is the corresponding morphism (\ref{3neu7}), we obtain the line bundles 
  \be
  T_i := \mathrm{Bun}(\mathcal{T}'_i \circ \mathcal{T}_i^{-1})
  \labl{3neu3}
over $M$, of curvature $\mathrm{curv}(T_i) \eq \rho_i' - \rho_i$. Then we have
  \be \bearll
  \widetilde{\mathcal{D}} \!\! &= (\mathcal{T}_2^{} \oti
  \id_{\mathcal{I}_{\bbf}}) \circ \mathcal{D} \circ \mathcal{T}_1^{-1}
  \\{}\\[-.7em] &
  \cong (\mathcal{T}_2^{} \circ (\mathcal{T}'_2)^{-1}\otimes
  \id_{\mathcal{I}_{\bbf}}) \circ (\mathcal{T}'_2 \oti
  \id_{\mathcal{I}_{\bbf}}) \circ \mathcal{D} \circ (\mathcal{T}'_1)^{-1}
  \circ \mathcal{T}'_1 \circ \mathcal{T}_1^{-1}
  \\{}\\[-.7em] &
  = (\mathcal{T}_2^{} \circ (\mathcal{T}'_2)^{-1}\otimes
  \id_{\mathcal{I}_{\bbf}}) \circ \widetilde{\mathcal{D}}' \circ
  \mathcal{T}'_1 \circ \mathcal{T}_1^{-1} .
  \eear \ee
Using the identification $\mathrm{Bun}$ of bundle gerbe morphisms with
vector bundles and its properties (\ref{3neu1}) and (\ref{4neu2}) we obtain
  \be
  E \cong T_2^{*} \otimes E' \otimes T_1 \,\text{.}
  \labl{3neu4}

\medskip

We can apply this result in the following way to the bi-brane $(Q,\mathcal{D})$.
The pullback of the bimodule $\mathcal{D}$ along the map $\phi_S{:}\ S \,{\to}\,
Q$ gives a 
$(\phi_1^{*}\mathcal{G}_1^{})|_S^{}$-$(\phi_2^{*}\mathcal{G}_2^{})|_S^{}$-bimodule
  \be
  \phi_S^{*}\mathcal{D}:\quad (\phi_1^{*}\mathcal{G}_1^{})|_S^{} \to
  (\phi_2^{*}\mathcal{G}_2^{})|_S^{} \otimes \mathcal{I}_{\phi_S^{*}\bbf}
  \,\text{.}
  \ee

The pullback bundle gerbes $\phi_i^{*}\mathcal{G}_i$ over $\Sigma_i$ are
trivializable by dimensional reasons. A choice $\mathcal{T}_i{:}\
\phi_i^{*}\mathcal{G}_i^{} \,{\to}\, \mathcal{I}_{\rho}$ of trivializations 
for two-forms $\rho_i$ on $\Sigma_i$ produces a vector bundle over $S$.
With this vector bundle $E$ we define
  \be
  \mathrm{hol}_{\mathcal{G}_1,\mathcal{G}_2,\mathcal{D}}(\Sigma,S)^{}
  := \exp \Big ( \mathrm{i}\! \int_{\Sigma_1}\! \rho_1 \Big ) \, \exp
  \Big ( \mathrm{i}\! \int_{\Sigma_2}\! \rho_2 \Big ) \,
  \mathrm{tr}(\mathrm{hol}_{E}^{}(S)) \in \C
  \labl{3neu5}
to be the holonomy in the presence of the bi-brane $(Q,\mathcal{E})$.
This holonomy is the appropriate generalization of the Wess\hy Zumino
\erf{neu1} term in situations where the simplifying assumptions
on the topology of the background and the bi-brane do not hold any longer.

This definition does not depend on the choice of the trivializations
$\mathcal{T}_1$ and $\mathcal{T}_2$, as we shall now establish. For different
choices $\mathcal{T}_1'$ and $\mathcal{T}_2'$ we obtain the line bundles $T_i$ 
introduced in (\ref{3neu3}). Since by construction we have $\partial \Sigma_1 
\eq S$ and $\partial \Sigma_2 \eq \overline{S}$, and since the curvature of the 
bundles $T_i$ is $\mathrm{curv}(T_i) \eq \rho_i' - \rho_i$, the holonomies 
of $T_1$ and $T_2$ around $S$ are given by
  \be
  \mathrm{hol}_{T_1}(S) = \exp\big( \mathrm{i}\!\int_{\Sigma_1}\!
  \rho'_1 \,{-}\, \rho_1 \big)
  \qquad\text{and}\qquad
  (\mathrm{hol}_{T_2}(S))^{-1} = \exp\big( \mathrm{i}\!\int_{\Sigma_2}\!
  \rho'_2 \,{-}\, \rho_2 \big) \,,
  \labl{5.25}
respectively. {}From (\ref{3neu4}) we obtain
  \be
  \mathrm{tr}(\mathrm{hol}_{E}(S)) =\mathrm{tr}(\mathrm{hol}_{ T_2^{*} \otimes
  E' \otimes T_1}(S)) =(\mathrm{hol}_{T_2}(S))^{-1} \,
  \mathrm{tr}(\mathrm{hol}_{E'}(S)) \, \mathrm{hol}_{T_1}(S) \,\text{.} 
  \ee
Together with \erf{5.25} this shows the independence
of number (\ref{3neu5}) of the choice of the trivializations.

\medskip

To discuss the relation between the holonomy (\ref{3neu5}) and the form of 
the Wess\hy Zumino term given in Section \ref{secWZ}, suppose there exist 
3-dimensional oriented submanifolds $B_1$ and $B_2$ in $M_1$ and $M_2$,
respectively, and a 2-dimensional oriented submanifold $D$ of $Q$ such that
  \be
  \partial D=\phi_S(S) \,\text{, }\quad~
  \partial B_1 = \phi_1(\Sigma_1) \,{\cup}\, p_1(\overline{D})
  \quad~\text{ and }\quad
  \partial B_2=\phi_2(\Sigma_2) \,{\cup}\, p_2(D)\,\text{.} 
  \labl{neu102}
By dimensional reasons we can choose trivializations
$\mathcal{T}_i{:}\ \mathcal{G}_i|_{\partial B_i}^{} \,{\to}\, 
\mathcal{I}_{\rho_i}$ of the two bundle gerbes over $\partial B_i$, 
thus producing a vector bundle $E$ over $D$ of curvature
  \be
  \Frac{1}{n}\,\mathrm{tr}(\mathrm{curv}(E))
  = \bbf|_D^{} + p_2^{*}\rho_2^{}|_D^{} - p_1^{*}\rho_1^{}|_D^{} \,\text{.}
  \labl{neu101}
The pullbacks $\phi_i^{*}\mathcal{T}_i{:}\ \phi_i^{*}\mathcal{G}_i \,{\to}\, 
\mathcal{I}_{\phi^{*}\rho_i}$ are trivializations as used in the definition of
the holonomy \erf{3neu5}, which hence becomes
  \be
  \mathrm{hol}_{\mathcal{G}_1,\mathcal{G}_2,\mathcal{D}}(\Sigma,S)
  = \exp \Big( \mathrm{i}\! \int_{\!\phi_1(\Sigma_1)}\! \rho_1 \Big) \, \exp
  \Big( \mathrm{i}\! \int_{\!\phi(\Sigma_2)}\! \rho_2 \Big) \,
  \mathrm{tr}(\mathrm{hol}_{E}(\phi_S(S))) \,\text{.}
  \ee
Here the holonomy of the vector bundle $E$ around the boundary $\phi_S(S)$ of
$D$ becomes by \erf{neu101}
  \be
  \mathrm{tr}(\mathrm{hol}_{E}(\phi_S(S)))
  = \mathrm{tr}(\mathrm{hol}_{E}(\partial D)) = \exp\big( \mathrm{i}\!\int_{\!D}
  \bbf+p_2^{*}\rho_2^{}-p_1^{*}\rho_1^{} \big) \,\text{.}
  \ee
The holonomy of the bundle gerbe  $\mathcal{G}_i|_{\partial B_i}^{}$ around 
the closed surface $\partial B_i$ is, by definition,
  \be
  \mathrm{hol}_{\mathcal{G}_i}(\partial B_i)
  = \exp\big(\mathrm{i}\!  \int_{\!\partial B_i} \rho_i^{} \big)
  = \exp\big(\mathrm{i}\! \int_{\!\phi_i(\Sigma_i)}\!
  \rho_i \,\pm\, \mathrm{i}\! \int_{\!D}\!p_i^{*}\rho_i^{} \big )
  \ee
with a minus sign for $i\eq1$ and a plus sign for $i\eq2$, according to the
relative orientations of $D$ and $\partial B_i$
in \erf{neu102}. On the other hand, we have
  \be
  \mathrm{hol}_{\mathcal{G}_i}(\partial B_i)
  = \exp\big(\mathrm{i}\! \int_{\!B_i}\! H_i^{} \big )
  \ee
with $H_i$ the curvature of $\mathcal{G}_i$. 
Taking the last four equalities together, we obtain
  \be
  \exp \Big (\mathrm{i}\! \int_{\!B_1}\! H_1 +\mathrm{i}\! \int_{\!B_2}\!
  H_2 +\mathrm{i}\! \int_{\!D}\! \bbf \Big)
  = \mathrm{hol}_{\mathcal{G}_1,\mathcal{G}_2,\mathcal{D}}(\Sigma,S) \,.
  \ee
We conclude that the holonomy of the bi-brane indeed specializes to the 
exponential of the Wess\hy Zumino term in the form given in Section \ref{secWZ}.

\newpage

\end{document}